%% file: main.tex
\documentclass[sigconf,natbib=true,nonacm]{acmart}

\input{preamble}

\setcopyright{none}
\renewcommand\footnotetextcopyrightpermission[1]{}

\begin{document}

\title{WeightRec: Evolving Sequential Recommenders with LLM-Guided Weight Edits}

\author{Xu Huang}
\email{xu\_huang@mail.ustc.edu.cn}
\affiliation{%
  \institution{University of Science and Technology of China}
  \city{Hefei}
  \country{China}}

\author{Jiawei Feng}
\email{jwf3ng@mail.ustc.edu.cn}
\affiliation{%
  \institution{University of Science and Technology of China}
  \city{Hefei}
  \country{China}}

\author{Hongzheng Qin}
\email{qinhz@mail.ustc.edu.cn}
\affiliation{%
  \institution{University of Science and Technology of China}
  \city{Hefei}
  \country{China}}

\author{Jiancan Wu}
\email{wujcan@gmail.com}
\affiliation{%
  \institution{University of Science and Technology of China}
  \city{Hefei}
  \country{China}}

\renewcommand{\shortauthors}{Huang et al.}

\input{chapters/0_abstract}



\maketitle

\input{chapters/teaser}

\input{chapters/1_introduction}
\input{chapters/framework_figure}
\input{chapters/2_preliminary}
\input{chapters/3_methodology}

\input{chapters/4_experiments}

\input{chapters/5_related_work}

\input{chapters/6_conclusion}
\label{main-text-end}

\bibliographystyle{ACM-Reference-Format}
\bibliography{references_updated}

\end{document}

%% file: preamble.tex
\usepackage{booktabs}
\usepackage{multirow}
\usepackage{algorithm}
\usepackage[noend]{algpseudocode}
\usepackage{xcolor}

\usepackage{pifont}
\usepackage{colortbl}
\usepackage{enumitem}
\usepackage{xspace}

\providecommand{\symbf}[1]{\mathbf{#1}}

\newcommand{\ie}{\textit{i.e.}\xspace}
\newcommand{\eg}{\textit{e.g.}\xspace}
\newcommand{\cf}{\textit{cf.}\xspace}

\newcommand{\I}{\mathcal{I}}

\newcommand{\D}{\mathcal{D}}
\newcommand{\Dfit}{\mathcal{D}^{\text{fit}}}
\newcommand{\Dval}{\mathcal{D}^{\text{val}}}
\newcommand{\T}{\mathcal{T}}
\newcommand{\E}{\symbf{E}}
\newcommand{\bb}{\symbf{b}}
\newcommand{\e}{\symbf{e}}
\newcommand{\h}{\symbf{h}}
\newcommand{\Mat}{\mathcal{M}}
\newcommand{\stat}{\sigma}
\newcommand{\ndcg}{\mathrm{NDCG}}
\newcommand{\chain}{c}

\definecolor{oursbg}{HTML}{EAF2FB}

\newcommand{\concept}[1]{#1}

%% file: chapters/0_abstract.tex
\begin{abstract}
 Sequential recommenders require continual adaptation as new user interactions arrive, item popularity changes, and previously unseen items enter the catalog. 
 Existing update methods typically rely on gradient-based fine-tuning or predefined post-hoc corrections, without explicitly diagnosing which model parameters contribute to performance degradation. 
 We propose WeightRec, an LLM-guided framework that updates trained sequential recommenders through direct weight editing. WeightRec enables an LLM agent to inspect model parameters, probe hidden representations, and evaluate interventions to identify promising parameter-level updates. 
 Each update is represented as a symbolic program that specifies the target parameters and computes edit values from feedback statistics. To reduce overfitting to the data used to construct edits, WeightRec evaluates candidate programs on chronologically subsequent interactions. 
 Accepted programs form an edit chain that can be re-materialized as new feedback arrives, either independently of gradient-based training or on top of a fine-tuned model. 
 Experiments on three datasets and three sequential recommendation backbones show that WeightRec achieves higher NDCG@10 than the evaluated post-hoc corrections in all nine single-update settings. 
 After three successive streaming updates, WeightRec improves NDCG@10 over incremental fine-tuning by 25\% on average as a standalone updater and by 28\% when combined with fine-tuning. These results demonstrate the potential of LLM-guided weight editing for adapting deployed recommender systems without modifying their inference architectures.

\end{abstract}

%% file: chapters/teaser.tex
\begin{figure*}[t]
  \centering
  \includegraphics[width=\textwidth]{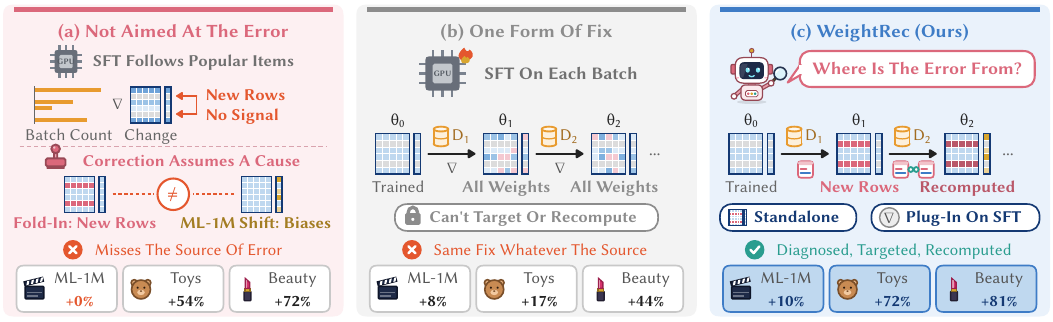}
  \vspace{-4mm}
  \caption{Three ways to update a trained recommender as feedback arrives. (a) Existing updates are not aimed at the source of error: the gradient of SFT follows the popular items of a batch and leaves new item rows without signal, and each post-hoc correction targets one cause fixed in advance, so it helps only where that cause dominates, \eg folding-in edits the rows of new items, whereas the shift on MovieLens-1M lies in the item biases. (b) They express only one form of fix: SFT applies a dense change of all weights on each batch $D_t$, whatever the source, which cannot be targeted or recomputed. (c) WeightRec lets an LLM locate the source of error and write the fix as an edit chain over the parameters it implicates, recomputed on each batch, alone or on top of SFT.}
  \Description{Three panels. The first shows that the gradient of fine-tuning on a batch changes the rows of popular items and not the rows of new items, and that a correction edits the rows of new items while the shift may lie in the biases. The second and third update a model over the same stream of feedback batches: fine-tuning applies the same dense change of all weights at each step, while WeightRec applies a growing chain of edit programs to the rows of new items and to biases, alone or on top of fine-tuning. Accuracy gains are shown below each panel.}
  \label{fig:teaser-figure}
\end{figure*}

%% file: chapters/1_introduction.tex
\section{Introduction}
\label{sec:introduction}
Sequential recommenders~\cite{GRU4Rec,SASRec,BERT4Rec} are trained on past interactions and then serve users whose behavior keeps changing, as new items arrive, the popularity of existing items drifts, and the interests of users move.
Therefore, a deployed recommender is updated as feedback streams in, and the standard update fine-tunes the model on each newly arrived batch of interactions by gradient descent (\ie supervised fine-tuning, SFT)~\cite{Rendle2008,SML}.
Training-free post-hoc corrections, such as prior shift~\cite{PriorShift,LogitAdjust}, popularity adjustment~\cite{PDA}, and folding-in of new items~\cite{FoldingIn}, offer a cheaper alternative for a single update.
Both families fix the form of an update before they see the feedback, and this paper asks whether an update can instead be written for what the feedback reveals about each model.

Existing update rules face two limitations.
First, they \textbf{do not aim the update at the source of error}, although what limits the model differs across datasets and backbones.
SFT lets the gradient of a small batch decide where the update goes, and the examples of a batch mostly target popular items.
As a result, the rows of new and rarely seen items, which often limit the model, receive limited positive supervision, while the rest of the model fits the batch~\cite{Kirkpatrick2017,SML}.
Each correction instead targets one cause fixed in advance (Figure~\ref{fig:teaser-figure}(a)), so it helps only where that cause dominates, \eg folding-in improves NDCG@10 considerably on Beauty and not at all on MovieLens-1M, where PDA is the better correction.
Aiming an update requires inspecting the model on the feedback, forming a hypothesis about the cause, and testing it, an open-ended reasoning process that no fixed rule performs.
Second, even when the source is known, they \textbf{can express only one form of fix}.
The fix a source calls for is a change to particular parameters, such as the rows of one group of items, a bias, or both, with values derived from the feedback.
Standard full-parameter SFT, however, determines weight changes through gradients on the feedback batch (Figure~\ref{fig:teaser-figure}(b)), without explicitly specifying the target parameters or representing the update as a reusable program, whereas each post-hoc correction applies a predefined adjustment.
Such parameter-level fixes form an open space that no single rule covers, and they are most naturally written as code over named parameters.
Today, both steps are left to an engineer who debugs the model and writes a fix by hand, which does not scale to every update of a stream.

To address these limitations, our key idea is to let an LLM agent take the role of this engineer, since LLMs can reason over the outputs of tools and write code.
Like an engineer under debugging, the agent inspects the state, states a hypothesis, intervenes on the suspected part, and observes how the output changes before committing a fix, an idea that causal tracing has made precise for neural networks~\cite{ROME}.
Each recommender offers the means for this loop, namely named parameters that can be read and written, intermediate states that can be recorded on the feedback, and item scores whose changes can be read on any group of examples.
The agent drives the loop through this interface and writes the update as a short program over named parameters that targets exactly the parameters its interventions implicate (Figure~\ref{fig:teaser-figure}(c)).
Since such a program computes its values from statistics of the feedback, it can be recomputed when a new batch arrives and applied on any weights, including the weights that SFT has just produced.

However, giving the agent this freedom poses a \textbf{challenge of judging an edit} without test users.
On the one hand, edits in a recommender are \concept{non-local}.
Since all items compete for the same top-$K$ lists, an edit that lifts the items a hypothesis names also pushes other items out of the lists of each user, and its gain on the targeted items says little about its effect on everyone else.
On the other hand, accuracy can be measured only on the feedback whose statistics the edits are built from, which leads to \concept{circular feedback}.
An edit that copies these statistics into the weights fits the very examples that score it, and it looks accurate on the feedback without carrying over to future users.

We present WeightRec, an update paradigm in which an LLM agent updates a trained recommender by \concept{weight editing} instead of gradient steps (Figure~\ref{fig:main-figure}).
Since WeightRec touches the model only through an interface of parameters, forward states, and item scores, the same procedure serves any backbone that exposes them.
Each update has three components, one for each limitation and one for the challenge.
(i) \textbf{Weight-space diagnosis} lets the agent read the parameters, probe hidden states, and intervene on the weights, so that it locates the source of error by testing hypotheses instead of assuming it in advance.
(ii) \textbf{Symbolic weight edits} express each update as a composition of operations whose row selectors are item features and whose values are computed by tools from statistics of the feedback, so that the form of each update follows the located source: one update can target a group of items, combine several operations, and be recomputed on new feedback.
(iii) \textbf{Forward-validated feedback} builds each statistic from earlier feedback and keeps an edit only if it gains on later feedback (termed \concept{forward validation}), which removes circular feedback by construction.
Accepted edits form a chain that each later update inherits and re-materializes with the newest statistics.
The chain is applied either to the trained model, so that WeightRec updates the model on its own, or to the model that SFT produces at the same step, so that WeightRec serves as a \concept{plug-in} for gradient updates.

We evaluate WeightRec on three datasets that differ in what changed after training and on three backbones with recurrent, causal, and bidirectional encoders, with test users who are disjoint from the feedback users and later in time.
First, a single update by WeightRec outperforms fine-tuning and published corrections tuned under the same protocol (\cf Section~\ref{sec:main-performance}).
Second, in streaming updates, both WeightRec alone and SFT followed by WeightRec outperform SFT in most settings and never fall significantly below it, improving NDCG@10 over SFT by about a quarter on average (\cf Section~\ref{sec:stream}).
Third, each component contributes, and forward validation is what keeps the agent from memorizing the feedback (\cf Section~\ref{sec:ablation}).
Finally, the edits follow what changed in each dataset, as a case study illustrates (\cf Section~\ref{sec:case}).
We highlight that WeightRec offers an update paradigm built on readable programs, which can update a recommender alone or extend an existing gradient pipeline without changing it.

%% file: chapters/framework_figure.tex
\begin{figure*}[t]
  \centering
  \includegraphics[width=\textwidth]{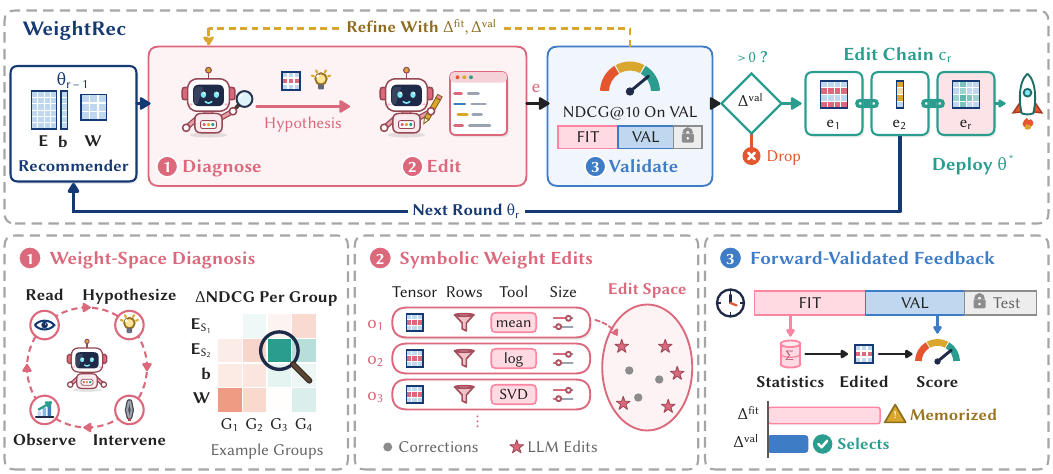}
  \vspace{-6mm}
  \caption{Overview of WeightRec. Top: in each round, the LLM diagnoses the model, writes a weight edit, and keeps the edit only if it improves accuracy on later feedback; kept edits form a chain that gives the next round's model and, after the last round, the deployed model. Bottom: \ding{172} diagnosis locates which weights limit which examples, \ding{173} edits are written as short symbolic operations whose values are computed by tools, and \ding{174} validation uses only feedback that comes after the data used to fit the edit.}
  \Description{One round of WeightRec: diagnose, edit, and validate, with a loop over rounds, and three panels detailing each component.}
  \label{fig:main-figure}
\end{figure*}

%% file: chapters/2_preliminary.tex
\section{Problem Formulation}
\label{sec:preliminary}

We first define the interface through which WeightRec accesses a recommender and its instance for sequential recommendation.
After that, we define the feedback that arrives after training, in one batch or as a stream, and formulate weight editing.
Finally, we review the two update rules in use, gradient fine-tuning and post-hoc correction.

\subsection{Recommender Interface}
\label{sec:recommender}

Let $\I=\{1,\dots,N\}$ be the item catalog and $s_u=(i_1,\dots,i_t)$ the interaction history of user $u$.
WeightRec treats a trained recommender $\theta=\{\symbf{W}_p\}_{p\in\mathcal{P}}$ as a set of named parameter tensors and assumes only three things of it.
(i) Every tensor can be read and written by name, and each is marked as editable or frozen.
(ii) The forward pass on a history $s_u$ can record the hidden state at a set of named sites, such as the output of each layer.
(iii) The model ranks the catalog by a score $s_\theta(u,i)$, so that the rank of any target item and the top-$K$ list of any user can be computed.
Nothing else about the architecture is assumed, and a new backbone enters WeightRec by a short description of its parameters, sites, and score.

In this paper, we instantiate the interface on sequential recommenders, whose score takes the form
\begin{equation}
  \label{eq:score}
  s_\theta(u,i) = \h_u^{\top}\e_i + b_i, \qquad \h_u = f_\Phi(s_u;\E),
\end{equation}
where $\e_i$ is row $i$ of the item embedding table $\E\in\mathbb{R}^{N\times d}$, $b_i$ is the item bias, and $f_\Phi$ is a sequence encoder with parameters $\Phi$ that reads the embeddings of the history items.
The table $\E$ is tied, meaning that the same row serves as the input embedding and the output vector of item $i$.
SASRec~\cite{SASRec}, GRU4Rec~\cite{GRU4Rec}, and BERT4Rec~\cite{BERT4Rec} all take this form and differ only in $f_\Phi$ (causal self-attention, a gated recurrent unit, and bidirectional self-attention over the history with a masked last position, respectively), and hence in their parameters and sites.
Recommendation ranks all items by $s_\theta(u,\cdot)$, excluding those already in $s_u$.

\subsection{Training, Feedback, and Test}
\label{sec:data}

The model $\theta_0$ is trained on all interactions before a time $T_0$.
After training, a log of feedback $\D$ becomes available, which contains the interactions of a set of feedback users in the window $[T_0,T_1)$.
Each example in $\D$ is a pair of a history and the next item the user chose.
In a \concept{streaming update}, $\D$ arrives in $S$ chronological batches $\D_1,\dots,\D_S$ and the model is updated after each batch, where the update at step $t$ may use only $\D_{1:t}=\D_1\cup\dots\cup\D_t$, and a single update is the case $S=1$.
Model quality is measured on a test set $\T$ of next-item examples at or after time $T_1$ from a disjoint set of users.
$\T$ is never visible during an update and is scored once for each deployed model.
Each item carries features from the item table of the dataset and from interaction counts, such as its training count, its count in the feedback, its genres, and whether it occurs before $T_0$.

\subsection{Weight Editing}
\label{sec:weight-editing}

Weight editing changes the trained weights directly, using only $\D$ and no gradient step,
\begin{equation}
  \label{eq:objective}
  \max_{\Delta}\ \ndcg@10\big(\theta_0+\Delta;\ \T\big), \qquad \Delta = \mathcal{A}(\theta_0,\D),
\end{equation}
where $\mathcal{A}$ is an editing procedure that may read $\theta_0$ and $\D$ but not $\T$.
$\mathcal{A}$ receives no description of how $\theta_0$ mismatches the behavior after $T_0$.
The loss may arise in any parameter, \eg rows that received little training signal, biases that follow training popularity, or encoder states that no longer match the interests of users, and which of them dominates differs across datasets and backbones.
$\mathcal{A}$ must therefore locate the mismatch from $\theta_0$ and $\D$ before it can remove it.
In a streaming update, $\mathcal{A}$ runs after each batch on the feedback $\D_{1:t}$ that has arrived, and the model deployed after the last batch is the one that matters.

\subsection{Gradient Updates and Post-Hoc Correction}
\label{sec:post-hoc}

The standard update rule is supervised fine-tuning (SFT), which trains all parameters on the newly arrived batch,
\begin{equation}
  \label{eq:sft}
  \theta_t = \mathrm{SFT}\big(\theta_{t-1};\ \D_t\big),
\end{equation}
where $\mathrm{SFT}$ minimizes the next-item cross-entropy of Equation~\ref{eq:score} on $\D_t$ by gradient descent with early stopping on held-out feedback users~\cite{Rendle2008,SML}.
Since most examples of a batch target popular items, the gradient of Equation~\ref{eq:sft} gives little signal to the rows of new and rare items.
The common training-free alternative is a post-hoc correction $g_\phi$ with one knob $\phi$ that is tuned on $\D$.
Prior shift adds $\tau(\log p_{\D}(i)-\log p_{\text{train}}(i))$ to each logit~\cite{PriorShift,LogitAdjust}.
PDA rescales scores by the recent popularity of the item raised to a power $\gamma$~\cite{PDA}.
Folding-in sets the row of a new item, which has no interaction before $T_0$, to the mean user state of the feedback users who chose it~\cite{FoldingIn}.
Both rules fix the form of the update before they see the feedback, SFT as a dense gradient step on all parameters and each correction as one adjustment along one direction.
Therefore, neither can restrict an update to the group of items that limits the model, combine several adjustments, or recompute an update when more feedback arrives.

%% file: chapters/3_methodology.tex
\section{WeightRec}
\label{sec:method}

WeightRec turns an LLM into an editor that updates a trained recommender through its weights (Figure~\ref{fig:main-figure}).
The LLM accesses the model only through the interface of Section~\ref{sec:recommender}, and every action below is defined on named parameters, recorded states, and item scores, so it applies unchanged to every backbone.
The three components of a round follow the order of a session.
We first describe how the LLM reads, probes, and intervenes on the model to locate what limits it, which answers the first limitation (Section~\ref{sec:diagnosis}).
After that, we define the symbolic edits it writes, which answer the second limitation (Section~\ref{sec:symbolic}).
Then, we show why an edit must be judged by its aggregate accuracy forward in time and how WeightRec does so, which answers the challenge of judging an edit (Section~\ref{sec:forward}).
We then put the steps together into one editing session (Section~\ref{sec:session}) and chain sessions into rounds (Section~\ref{sec:evolve}).
Finally, we turn the rounds into update steps of a stream, with WeightRec on its own or as a plug-in for SFT (Section~\ref{sec:stream-method}).

\subsection{Weight-Space Diagnosis}
\label{sec:diagnosis}

To let the LLM locate what limits the model without being told, WeightRec exposes the model as an object to debug, through three types of actions that follow the inspect, intervene, and observe loop of causal tracing~\cite{ROME} (step \ding{172} of Figure~\ref{fig:main-figure} and Table~\ref{tab:actions}).
The LLM works in turns, and each turn returns one action with a short rationale, in the style of ReAct~\cite{ReAct}.

\begin{table}[t]
  \centering
  \caption{Diagnosis actions. Each is defined on the interface of Section~\ref{sec:recommender} and is shared by all backbones.}
  \label{tab:actions}
  \vspace{-3mm}
  \small
  \setlength{\tabcolsep}{3.0pt}
  \begin{tabular}{@{}l|p{0.30\columnwidth}|p{0.47\columnwidth}@{}}
    \toprule
    \textbf{Type} & \textbf{Acts on} & \textbf{Returns} \\
    \midrule
    Read & parameters, item features & shapes, norms, singular values, statistics of item-indexed rows by item group, nearest items \\
    Probe & $\theta$ on FIT examples & hidden states at a named site, accuracy per group, recommendation audit, examples with top-$K$ lists \\
    Intervene & a copy of $\theta$ with a candidate edit & accuracy change on FIT and VAL per group, and every probe repeated on the edited copy \\
    \bottomrule
  \end{tabular}
  \vspace{-3mm}
\end{table}

\textbf{Read.}
Read actions inspect the parameters without running the model, namely the shape and spectrum of any tensor, the statistics of the rows of item-indexed tensors grouped by an item feature and their correlation with item features, and the nearest items of any vector.

\textbf{Probe.}
Probes and interventions report accuracy per \concept{group} of feedback examples, which are grouped by the target item or the user, \eg popularity groups of the target or groups of users by history length.
Probe actions run the model on feedback examples and record what it computes, \eg the hidden states at a named site for the users whose next item matches a selector, the feedback accuracy per group, a recommendation audit of the items that are recommended far more or far less often than feedback users choose them, and a few examples with their top recommendations.
Each session starts with an initial report computed by these read and probe actions on the unedited model.

\textbf{Intervene.}
An intervention applies a candidate edit to a copy of the weights and reports how the predictions changed, namely the accuracy change per group and every probe repeated on the edited copy, such as the rank of the target in individual examples and the items that the edited copy recommends far more or far less often than they are chosen.
Comparing the probes before and after the intervention shows which groups of FIT examples respond to the edited parameters and which items are displaced elsewhere, so the LLM can confirm or reject a hypothesis (\eg ``these rows are too weak to compete for these users'') and revise it before it commits.
Interventions serve to locate, not to decide.
Since an edit of any parameter moves the top-$K$ lists of all users, a gain on the group a hypothesis names says little about the aggregate, and whether an edit is kept is decided only by its forward-validated aggregate gain (Section~\ref{sec:forward}).
Since the LLM decides where to edit from what it reads and from how the model responds to its interventions, it needs no prior description of the mismatch, and this loop also distinguishes WeightRec from a search over edits.

\subsection{Symbolic Weight Edits}
\label{sec:symbolic}

To let one edit act on exactly the rows a hypothesis names, WeightRec lets the LLM write each edit as a program (step \ding{173}).
Specifically, an edit $e=(o_1,\dots,o_K)$ is a sequence of operations, and each operation names a parameter of the interface, a row selector, a value, and a strength.
The value of an operation is not typed in by the LLM but computed by a tool that it chooses from a small library (\eg a mean of user states, a log-count transform, or a singular direction), so that the LLM decides which algorithm computes how the selected rows change.
Operations come in a few generic types, namely setting, adding to, or scaling rows of any tensor, and low-rank updates and direction ablations of any matrix, so that no operation refers to a specific architecture.
For item-indexed tensors such as $\E$ and $\bb$, a selector is a predicate over item features, such as ``new items'', ``training count below 10'', or a genre.
A value is either a number or a vector expression over statistics and recorded states, such as the mean embedding of trained items in the same genre, the mean hidden state of the feedback users who chose the item, or a stored direction.
A bias operation may also take a feature-valued form,
\begin{equation}
  \label{eq:bias-op}
  b_i \leftarrow b_i + \alpha\, \psi\big(x_i\big), \qquad i \in \mathrm{SEL},
\end{equation}
where $x_i$ is an item feature (\eg its count in the feedback), $\psi$ is a transform such as $\log(1+\cdot)$ or a z-score, and $\alpha$ is a strength.
Rank-one updates $\symbf{W}\leftarrow\symbf{W}+s\,\symbf{u}\symbf{v}^{\top}$ and direction ablations reach every editable matrix of the encoder $f_\Phi$.
An edit becomes weights through materialization,
\begin{equation}
  \label{eq:materialize}
  \theta' = \Mat\big(e;\ \theta,\ \stat(\mathcal{S})\big),
\end{equation}
where $\stat(\mathcal{S})$ collects the statistics an edit may use (item counts, user-state means, stored hidden states), computed on a data split $\mathcal{S}$.
Equation~\ref{eq:materialize} separates what an edit does from the data it is computed on.
As a result, the same edit can be materialized on one split for selection and on another for deployment, several operations compose into one edit, and the LLM can read back any edit it wrote.
To calibrate a strength such as $\alpha$, the LLM may replace one number with up to four values, and each value is evaluated as its own candidate.

\textbf{Expressiveness.}
The edit language contains each post-hoc correction of Section~\ref{sec:post-hoc} as a single operation, \eg prior shift is a feature-valued bias operation (Equation~\ref{eq:bias-op}) and folding-in is a row operation on new items whose value is a mean of user states.
Each correction is therefore one point in the space WeightRec explores, and the LLM can restrict it to the items where it helps, combine it with others, or replace it with edits that no single family expresses.

\subsection{Forward-Validated Feedback}
\label{sec:forward}

To judge an edit by the gain it brings to future users, WeightRec scores every edit by its aggregate accuracy, because edits are non-local, and never scores it on the interactions its statistics came from.
The need for this follows from Equation~\ref{eq:score}.
Consider folding-in applied to a trained item $i$, so that $\e_i$ becomes the mean state of the $n_i$ examples in $\D$ whose next item is $i$.
For one of these examples $(u,i)$, the new score is
\begin{equation}
  \label{eq:circular}
  \h_u^{\top}\e_i = \frac{1}{n_i}\lVert\h_u\rVert^2 + \frac{1}{n_i}\sum_{u'\neq u}\h_u^{\top}\h_{u'},
\end{equation}
where the first term (termed the self-match) exists only because the scored example was counted in the statistic.
A future user who chooses $i$ never receives this term.
The self-match grows as $n_i$ shrinks, so an editor that grades edits on $\D$ is pushed toward rewriting the rows of rare items with the answers of the very examples that grade them.
Count-based statistics in Equation~\ref{eq:bias-op} carry the same bias in a milder form.

WeightRec removes this circular feedback with forward validation (step \ding{174}).
Specifically, it splits $\D$ at the median time $M$ of its next-item events.
The earlier part $\Dfit$ (examples with targets before $M$, and all events before $M$) is the only source of statistics, of probes, and of audits.
The later part $\Dval$ is used only to score candidates.
For a candidate edit $e$ on a model $\theta$, the LLM receives
\begin{equation}
  \label{eq:val-gain}
  \Delta^{\text{val}}(e) = \ndcg\big(\Mat(e;\theta,\stat(\Dfit));\ \Dval\big) - \ndcg\big(\theta;\ \Dval\big),
\end{equation}
together with its paired 95\% confidence interval, the analogous gain $\Delta^{\text{fit}}(e)$ on $\Dfit$, and the gap $\Delta^{\text{fit}}-\Delta^{\text{val}}$, which return to the LLM (the dashed loop in Figure~\ref{fig:main-figure}).
A large gap signals memorization, and the LLM is instructed to select by $\Delta^{\text{val}}$ only.
The split runs forward in time, matching deployment, where the model is applied to interactions after the ones it was edited on.

\textbf{Re-materialization on all feedback.}
Selection uses only FIT statistics, but deployment can use all of them.
Once editing ends, WeightRec materializes the selected program again with statistics recomputed on all feedback (Equation~\ref{eq:deploy} in Section~\ref{sec:evolve}).
This step involves no training.
The selectors, strengths, and norms that the LLM chose stay fixed, and only the statistics inside the edit are recounted and re-averaged, \eg the feedback count in $\alpha\log(1+x_i)$ is counted over all of $\D$ instead of $\Dfit$.
It is safe because test interactions occur after $T_1$ and never enter $\stat(\D)$, and it is useful because statistics such as the counts of new items roughly double.
A fixed vector of edited weights could not be recomputed in this way, which is why the program form of Section~\ref{sec:symbolic} matters for deployment.

\subsection{The Editing Session}
\label{sec:session}

A session edits one starting model $\theta$ and runs until the LLM submits a trial, and if the LLM runs out of turns, the trial with the best VAL gain is submitted.
Each evaluation is one materialization on $\Dfit$ and one scoring on $\Dfit$ and $\Dval$, without any gradient step.
Every intervention of Section~\ref{sec:diagnosis} is such an evaluation, so the same trial serves both to test a hypothesis and as a candidate for submission.

\textbf{Budget by ideas.}
WeightRec gives the LLM a budget of edit ideas $K$, where one proposal counts as one idea whether it is a single edit or a sweep over up to four values, and caps the total number of evaluations at $B$.
A budget that counts every evaluation invites the LLM to spend most of it on sweeping the strength of its first idea, which leaves the other hypotheses untested.
Counting ideas instead lets the LLM test several hypotheses (different parameters, item groups, and mechanisms) before it refines the one that transfers, as the instructions to the LLM ask.

\subsection{Self-Evolution over Rounds}
\label{sec:evolve}

To let the recommender keep improving after its first edit, WeightRec repeats the session over rounds and lets each round build on the accepted edits.
Specifically, round $r$ starts from the edit chain $\chain_{r-1}=(e_1,\dots,e_{r-1})$ of all accepted edits, materialized on $\Dfit$ for selection,
\begin{equation}
  \label{eq:chain}
  \theta_{r-1} = \Mat\big(\chain_{r-1};\ \theta_0,\ \stat(\Dfit)\big),
\end{equation}
where materializing a chain applies its edits in order, each with statistics recomputed on the weights the previous edit produced.
The LLM runs a full session on $\theta_{r-1}$, so its reads, probes, and interventions act on the edited weights, and limitations that the earlier edits exposed become visible.
For example, once an earlier edit changes which items compete for the top of the lists, the biases and encoder matrices that score these items decide whether the change holds, and a later round can edit them.
The submitted edit $e_r$ is accepted if its gain over $\theta_{r-1}$ on $\Dval$ is positive (the gate in Figure~\ref{fig:main-figure}), giving $\chain_r=\chain_{r-1}\circ e_r$, and it is dropped otherwise.

Two records reach the LLM in every round.
The history $\mathcal{H}_r$ lists, for every earlier round, the submitted edit, its rationale, its VAL gain, and whether it was accepted, so the LLM neither repeats a rejected idea nor undoes an accepted one.
The experience $\mathcal{K}$ is a small library of generic weight recipes (a feedback-popularity prior on biases, a shift from training to feedback counts, and two ways to initialize the rows of new items), which are evaluated on $\theta_{r-1}$ at the start of every round and offered to the LLM as optional starting points.
The LLM decides whether to use, extend, or ignore them.
Since every link of the chain is a symbolic program, the chain is never a stack of stale numbers, and after the last round $R$ it is re-materialized on all feedback and deployed once,
\begin{equation}
  \label{eq:deploy}
  \theta^{*} = \Mat\big(\chain_R;\ \theta_0,\ \stat(\D)\big).
\end{equation}

\textbf{Lite and Evo.}
We call WeightRec-Evo the procedure with $R=3$ rounds, and WeightRec-Lite a single session on $\theta_0$ without history or experience, which is the early-stopped form of the procedure.

\textbf{Deployment.}
The deployed model keeps the architecture and the inference cost of $\theta_0$, since each operation changes parameter values and none adds a module, and each change is auditable and reversible, since dropping a link and re-materializing restores the earlier model.

\subsection{Streaming Updates via Edit Chains}
\label{sec:stream-method}

To update a recommender as feedback arrives, WeightRec turns the rounds of Section~\ref{sec:evolve} into update steps.
Specifically, round $t$ runs when batch $\D_t$ arrives, its FIT part holds all earlier batches and the earlier half of $\D_t$, and its VAL part is the later half of $\D_t$, thereby validating each step on the newest feedback.
The accepted edits form the chain $\chain_t$, and at each step the whole chain is re-materialized with the statistics of all arrived feedback $\D_{1:t}$.
WeightRec uses this chain in two ways.

\textbf{Standalone update.}
WeightRec updates the trained model by the chain alone,
\begin{equation}
  \label{eq:stream-alone}
  \theta_t = \Mat\big(\chain_t;\ \theta_0,\ \stat(\D_{1:t})\big),
\end{equation}
where no gradient step is taken and each earlier edit is recomputed on the newest counts and user states, \eg a feedback-count prior counts the new batch as soon as it arrives.

\textbf{Plug-in for SFT.}
WeightRec can also follow an existing gradient update.
The SFT trajectory $\mathcal{S}_t=\mathrm{SFT}(\mathcal{S}_{t-1};\D_t)$ of Equation~\ref{eq:sft}, with $\mathcal{S}_0=\theta_0$, is left as it is, and the chain is applied on top of it,
\begin{equation}
  \label{eq:stream-plug}
  \theta_t = \Mat\big(\chain_t;\ \mathcal{S}_t,\ \stat(\D_{1:t})\big).
\end{equation}
The edit of step $t$ is selected on $\mathrm{SFT}(\mathcal{S}_{t-1})$ fitted on the earlier half of $\D_t$, which keeps VAL out of each gradient, and it is re-materialized on $\mathcal{S}_t$ for deployment.
Since SFT never trains on edited weights, later gradient steps do not overwrite the edits, each step fine-tunes first and edits last, and dropping the chain returns the SFT model exactly.
As a result, WeightRec can be added to an existing update pipeline without changing it.

%% file: chapters/4_experiments.tex
\section{Experiments}
\label{sec:experiments}

\input{chapters/tab_main}

We evaluate WeightRec through four research questions.
\begin{itemize}[leftmargin=*,nosep]
  \item \textbf{RQ1}: Does a single update by WeightRec improve trained recommenders more than fine-tuning and post-hoc corrections?
  \item \textbf{RQ2}: In streaming updates, does WeightRec outperform SFT on its own and as a plug-in after SFT?
  \item \textbf{RQ3}: What do self-evolution, forward validation, and LLM diagnosis each contribute?
  \item \textbf{RQ4}: What do the edits do, and what does the LLM read and write to make them?
\end{itemize}

\subsection{Experiment Setup}
\label{sec:setup}

\textbf{Datasets.}
We use three datasets that differ in what changed after training, namely Beauty and Toys and Games (Toys) from Amazon reviews~\cite{AmazonData,Amazon2018} (5-core), and MovieLens-1M~\cite{MovieLens}.
New items make up 13\% of the test targets on Beauty, 10\% on Toys, and none on MovieLens-1M, where every target was trained but preferences have moved, and no method is told which of these changes matters.
For each dataset, $T_0$ and $T_1$ are the 80\% and 90\% quantiles of all timestamps, and users alternate between the feedback group and the test group, so feedback users and test users never overlap.
Each user contributes up to five next-item examples after $T_0$, the feedback examples lie in $[T_0,T_1)$, and the test examples lie at or after $T_1$, following recommendations against temporal leakage~\cite{RecLeakage}.
Table~\ref{tab:datasets} summarizes them.

\begin{table}[t]
  \centering
  \caption{Dataset statistics. Items counts the catalog of the model. ``New'' is the share of test targets without any training interaction.}
  \label{tab:datasets}
  \vspace{-3mm}
  \small
  \setlength{\tabcolsep}{3.0pt}
  \begin{tabular}{@{}l|rrrrrr@{}}
    \toprule
    \textbf{Dataset} & \textbf{Users} & \textbf{Items} & \textbf{Inter.} & \textbf{Feedb.} & \textbf{Test} & \textbf{New} \\
    \midrule
    Beauty & 22,363 & 11,937 & 198.5K & 5,754 & 5,094 & 13.2\% \\
    Toys & 19,412 & 11,770 & 167.6K & 4,368 & 4,284 & 9.8\% \\
    MovieLens-1M & 6,040 & 3,670 & 1.0M & 2,474 & 2,839 & 0.0\% \\
    \bottomrule
  \end{tabular}
  \vspace{-3mm}
\end{table}

\textbf{Backbones.}
We train SASRec~\cite{SASRec} (two blocks), GRU4Rec~\cite{GRU4Rec} (one layer), and BERT4Rec~\cite{BERT4Rec} (two bidirectional blocks), all with 64-dimensional tied item embeddings and an item bias, on the interactions before $T_0$ with early stopping, with three seeds each.
The three backbones share the actions, the edit language, and the instructions of WeightRec, and only the short description of the parameters, forward sites, and score in the prompt differs.

\textbf{Baselines.}
Fine-tuning updates all parameters on the feedback examples with early stopping on a held-out 20\% of feedback users, and serves as a reference for what the feedback can buy with training.
The training-free corrections are prior shift~\cite{PriorShift,LogitAdjust}, PDA~\cite{PDA}, folding-in of new items~\cite{FoldingIn}, and genre-mean initialization of new items, and Best correction picks, per run, the one with the highest VAL score.
For fairness, every correction computes its statistics on $\Dfit$, selects its knob on $\Dval$, and is re-materialized with statistics from $\D$, exactly as WeightRec does.

\textbf{Evaluation and implementation.}
We report HR and NDCG at cutoffs 3 and 10 over the full catalog~\cite{SampledMetrics}, excluding history items, and compare methods with paired bootstrap 95\% confidence intervals over the test examples of the three seeds.
The editor is the open-weight LLM hy4, prompted with a description of the backbone, the actions it may take, and working rules that tell it to judge an edit by its VAL gain only, and the task message never names a cause.
Each round allows $K=20$ edit ideas with at most $B=80$ evaluations.

\textbf{Streaming updates.}
For RQ2, the feedback of each dataset arrives in $S=3$ chronological batches of equal size, each method updates the model after each batch, and $\T$ scores the model deployed after the last batch.
Incremental SFT (SFT for short) applies Equation~\ref{eq:sft} with the hyperparameters of fine-tuning to the model of the previous step, while Cumulative SFT restarts from $\theta_0$ at each step and fine-tunes on all arrived feedback.
WeightRec runs one session per step with the history of earlier steps and the same budget as above, and the plug-in shares the trajectory of SFT, thereby differing from SFT only by the edits.

\subsection{(RQ1) A Single Update}
\label{sec:main-performance}

Table~\ref{tab:main-table} reports test accuracy after a single update on the nine combinations of dataset and backbone.
We make the following observations.

\textbf{WeightRec beats every correction in every setting.}
WeightRec improves the unedited model in all nine settings, and its NDCG@10 is above every correction in all nine, raising NDCG@10 over the unedited model by 60\% on average against 19\% for the best correction and 29\% for fine-tuning.
The gains are largest where the best correction picks the wrong family, such as Beauty with GRU4Rec and Toys with BERT4Rec.
Since WeightRec gains on datasets that differ in what changed after training and on three different encoders with the same actions and edit language, the LLM locates the edit each setting needs, which Section~\ref{sec:case} examines.

\textbf{No single correction is best across datasets.}
Folding-in is the strongest correction on Beauty and Toys, where new items are among the test targets, but it leaves MovieLens-1M unchanged, whereas PDA is the strongest correction on MovieLens-1M.
Therefore, a correction must match a cause that the practitioner rarely knows in advance.

\textbf{WeightRec exceeds fine-tuning where items have little training data.}
On the two Amazon datasets, WeightRec-Evo exceeds fine-tuning in all six settings, where a few thousand feedback examples are too few to fit the rows of new and rarely seen items by gradients.

\subsection{(RQ2) Streaming Updates}
\label{sec:stream}

\input{chapters/tab_stream}

Given that a single update by WeightRec exceeds fine-tuning, we next test whether this holds when the model is updated after each batch of a stream.
Table~\ref{tab:stream} reports test accuracy after three update steps, where WeightRec updates the model on its own (Equation~\ref{eq:stream-alone}) or as a plug-in after SFT (Equation~\ref{eq:stream-plug}).
We have the following observations.

\textbf{WeightRec alone outperforms SFT as a streaming update.}
WeightRec is significantly above SFT in NDCG@10 in seven of the nine settings and never significantly below it, and it improves NDCG@10 over SFT by 25\% on average, compared with 8\% for fine-tuning on all arrived feedback.
The gap is largest on Toys, where SFT barely improves over no update (1.42 compared with 1.33 with SASRec) while WeightRec reaches 2.12.
We attribute this gap to the items that appear in later batches, whose rows a small batch cannot fit by gradients, while the edit chain re-materializes their rows and biases from the statistics of each new batch.

\textbf{As a plug-in, WeightRec adds accuracy on top of SFT.}
SFT followed by WeightRec is significantly above SFT in six of the nine settings and never significantly below it, improving NDCG@10 over SFT by 28\% on average and by up to two thirds on Beauty with BERT4Rec (1.72 to 2.85).
It also significantly exceeds fine-tuning on all arrived feedback in six settings.
Since the SFT trajectory is the same with and without the plug-in, this gain comes from the edits alone, which suggests that the edits reach the parameters that gradient steps on a small batch leave under-fitted.
Such results show that WeightRec can update a recommender by itself or improve an existing gradient update, and that in seven of the nine settings at least one of the two uses is significantly above SFT.

\textbf{Re-materialized chains and LLM diagnosis both matter in the stream.}
Writing each edit into the weights and fine-tuning the edited weights at the next step is never significantly better than the plug-in, which is significantly more accurate in two settings, by 0.73 points on Beauty with BERT4Rec and 0.50 on Toys with GRU4Rec.
We attribute this difference to later gradient steps, which overwrite edits that are written into the weights.
Replacing the LLM in the plug-in by the fixed editor of Section~\ref{sec:ablation} reduces NDCG@10 significantly in five settings, which indicates that the stream also needs the edits that the LLM locates.

\subsection{(RQ3) Ablation Study}
\label{sec:ablation}

\input{chapters/tab_ablation}

To isolate each component, Table~\ref{tab:ablation} studies Beauty and MovieLens-1M with GRU4Rec, one setting limited by new items and one by drifted popularity.
The variants below WeightRec-Lite change its single session.
Without LLM (fixed editor), the best correction is extended in each of three rounds by the candidate with the largest VAL gain among the re-tuned corrections and 80 random edits of the same language.
Without forward validation, the LLM computes statistics and scores on all of $\D$.
Without re-materialization, the edit is deployed with its FIT statistics.
Without the idea budget, every evaluation counts against a budget of 20.
Without LLM, random edits of the same language are scored by the same rule, with 20 or 100 evaluations.
We make the following observations.

\textbf{Self-evolution refines what the first round starts.}
In the 27 WeightRec-Evo runs, 80 of the 81 submitted edits gain on VAL and join the chain.
The first round makes the broad change, mostly a feedback prior on the biases and new rows set from user states, and the later rounds start from these edited weights and recalibrate the biases of new items or reach the sequence encoder, because finer limitations become visible only in the edited weights.
The test gain over the unedited model grows from 0.84 to 0.91 NDCG@10 points over the rounds, so the first round carries most of the gain and WeightRec-Lite is its early-stopped form.

\textbf{Each design choice contributes.}
The fixed editor exceeds the best single correction in all nine settings, yet WeightRec-Evo, which is offered the same corrections, is significantly above it in seven, so the gain comes from what the LLM edits beyond stacking them.
Without re-materialization, the same session loses accuracy in both settings, since the statistics behind each edit double when recomputed on all feedback, which improves 25 of the 27 edits at no selection cost.
Counting ideas instead of evaluations lets the LLM test 19.1 ideas per session instead of 5.7, and both random searches fall below WeightRec-Lite although they share the edit language and the selection rule, so the difference comes from where the LLM chooses to edit after reading the weights.

\textbf{Forward validation keeps the LLM from memorizing the feedback.}
Without forward validation, the same session falls below the unedited model on MovieLens-1M.
Running WeightRec-Lite under both rules in all nine settings shows why.
Under the circular rule, the submitted edits gain 13 times more on the feedback they are scored on than on test, and in 9 of the 27 sessions the LLM writes the mean feedback-user state into the rows of trained items, which confirms the self-match term of Equation~\ref{eq:circular}.
Under forward validation, only 2 of the 27 submissions do so, while 10 restrict this statistic to new items, and forward validation is more accurate than the circular rule in seven of the nine settings.
\label{sec:feedback}

\subsection{(RQ4) What the Edits Do}
\label{sec:case}

To see why the updates help, we finally examine what the LLM writes in the single-update sessions.

\textbf{The edits follow what changed.}
The LLM is never told what limits the model, yet in the nine WeightRec-Lite sessions on each dataset it rewrites the rows of new items in seven on Beauty, where new items are 13\% of the targets, in six on Toys (10\%), and in none on MovieLens-1M, where it instead recalibrates the existing biases in all nine.
The unedited model never ranks a new item in the top 10, and WeightRec raises HR@10 on new-item targets to 15\% to 28\%, above the best correction in all six settings of Beauty and Toys.
These are the rows that a small batch of feedback fits poorly by gradients, which is consistent with the gains of the plug-in over SFT in Section~\ref{sec:stream}.

\begin{figure}[t]
  \centering
  \includegraphics[width=\columnwidth]{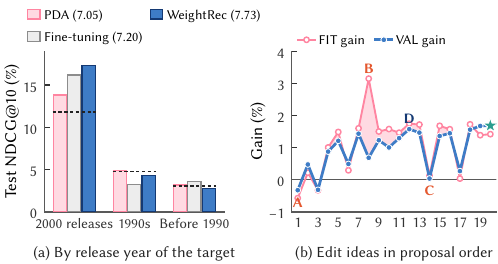}
  \vspace{-6mm}
  \caption{One WeightRec-Lite session (MovieLens-1M, SASRec). (a) Test NDCG@10 by the release year of the target, with the overall NDCG@10 in the legend and the unedited model dashed. (b) FIT and VAL gains of the 20 edit ideas, each at its trial with the best VAL gain: (A) shrinking the training-time biases, (B) moving item rows toward the states of their FIT users, (C) ablating a state direction in the attention values, (D) a bias for releases from 2000 on top of a feedback-count prior. The star marks the submitted edit.}
  \Description{Grouped bars of test NDCG@10 for targets released in 2000, in the 1990s, and before 1990 for PDA, fine-tuning, and WeightRec, and two lines of FIT and VAL gains over 20 edit ideas.}
  \label{fig:case}
\end{figure}

\textbf{Discovering limitations beyond predefined corrections.}
We follow one session on MovieLens-1M with SASRec, where no target is a new item (Figure~\ref{fig:case}).
The LLM first probes the states of the FIT users whose next movie was rarely seen in training and reads the nearest items of their mean state, many of which are films released in 2000 (\eg \emph{Almost Famous}, \emph{High Fidelity}).
It concludes that the encoder already represents the taste of these users for recent releases, while their few training interactions left the biases of these films low, a cohort that neither prior shift nor PDA can express.
The LLM tests the alternatives before it commits (Figure~\ref{fig:case}(b)), rejecting a bias shrinkage that loses on VAL (A), row rewrites that gain on FIT but not on VAL (B), and a direction ablation that leaves VAL unchanged (C).
It keeps a feedback-count prior with an additional bias for films released from 2000 (D), which raises their test NDCG@10 from 11.8 to 17.3, against 13.8 for PDA and 16.2 for fine-tuning.
The submitted edit is the following program, and a selector on the release year appears in most edits on MovieLens-1M, although the task message never mentions it.
\begin{center}
\small
\begin{tabular}{@{}l@{}}
\texttt{add\_bias(rows=all, 1.85 * zscore(log1p(feedback\_count)))} \\
\texttt{add\_bias(rows=\{year >= 2000\}, 1.2)}
\end{tabular}
\end{center}

%% file: chapters/tab_main.tex
\begin{table*}[t]
  \centering
  \caption{Test accuracy (\%) of the edited recommenders, averaged over three seeds. H and N denote HR and NDCG. All corrections select their knob by forward validation and are re-materialized with statistics from all feedback, the same protocol as WeightRec. Best correction is the correction that VAL selects in each run. Within each backbone, the best result in each column is in bold and the second best is underlined. $^{*}$The paired 95\% confidence interval of WeightRec minus the best correction lies above zero. $^{\dagger}$Uses gradient steps.}
  \label{tab:main-table}
  \vspace{-3mm}
  \small
  \setlength{\tabcolsep}{3pt}
  \begin{tabular*}{\textwidth}{@{\extracolsep{\fill}}ll|cccc|cccc|cccc@{}}
    \toprule
    & & \multicolumn{4}{c|}{\textbf{Beauty}} & \multicolumn{4}{c|}{\textbf{Toys}} & \multicolumn{4}{c}{\textbf{MovieLens-1M}} \\
    \textbf{Backbone} & \textbf{Method} & H@3 & N@3 & H@10 & N@10 & H@3 & N@3 & H@10 & N@10 & H@3 & N@3 & H@10 & N@10 \\
    \midrule
    \multirow{8}{*}{\rotatebox{90}{SASRec}} & Unedited & 1.16 & 0.86 & 2.51 & 1.32 & 1.14 & 0.85 & 2.50 & 1.33 & 5.51 & 4.04 & 12.97 & 6.63 \\
     & Fine-tuning$^{\dagger}$ & 1.53 & 1.13 & 3.42 & 1.80 & 1.15 & 0.90 & 2.67 & 1.43 & 5.42 & 4.03 & \textbf{14.84} & \underline{7.32} \\
     & Prior shift & 0.84 & 0.62 & 1.96 & 1.01 & 0.61 & 0.45 & 2.19 & 0.98 & 5.25 & 3.99 & 12.82 & 6.64 \\
     & PDA & 1.20 & 0.89 & 2.91 & 1.48 & 1.39 & \underline{1.13} & 2.85 & 1.63 & \underline{6.05} & \underline{4.44} & 13.74 & 7.12 \\
     & Folding-in & \textbf{1.82} & \textbf{1.36} & \underline{4.29} & \underline{2.22} & \underline{1.52} & \underline{1.13} & \underline{3.91} & \underline{1.96} & 5.51 & 4.04 & 12.97 & 6.63 \\
     & Genre init. & 1.15 & 0.83 & 2.44 & 1.28 & 1.14 & 0.85 & 2.53 & 1.34 & 5.51 & 4.04 & 12.97 & 6.63 \\
     & Best correction & 1.62 & 1.18 & 3.80 & 1.95 & 1.38 & 1.06 & 3.51 & 1.80 & \underline{6.05} & \underline{4.44} & 13.74 & 7.12 \\
     & \textbf{WeightRec (Ours)} & \underline{1.81} & \underline{1.31} & \textbf{4.75}$^{*}$ & \textbf{2.33}$^{*}$ & \textbf{1.53} & \textbf{1.14} & \textbf{4.15}$^{*}$ & \textbf{2.04}$^{*}$ & \textbf{6.26} & \textbf{4.69}$^{*}$ & \underline{14.32}$^{*}$ & \textbf{7.52}$^{*}$ \\
    \midrule
    \multirow{8}{*}{\rotatebox{90}{GRU4Rec}} & Unedited & 0.92 & 0.70 & 1.95 & 1.07 & 0.81 & 0.62 & 1.76 & 0.95 & \underline{5.35} & 3.94 & 11.69 & 6.16 \\
     & Fine-tuning$^{\dagger}$ & \underline{1.77} & \underline{1.30} & \underline{4.57} & \underline{2.29} & 0.82 & 0.65 & 2.25 & 1.15 & 5.15 & 3.84 & \underline{12.48} & \underline{6.41} \\
     & Prior shift & 1.04 & 0.81 & 2.24 & 1.23 & 0.79 & 0.58 & 1.77 & 0.94 & 5.34 & \underline{3.98} & 12.11 & 6.35 \\
     & PDA & 0.84 & 0.61 & 2.15 & 1.07 & 0.98 & 0.75 & 2.32 & 1.22 & 5.18 & 3.82 & 11.99 & 6.21 \\
     & Folding-in & 1.58 & 1.11 & \textbf{4.62} & 2.16 & \underline{1.13} & \underline{0.80} & \underline{3.42} & \underline{1.58} & \underline{5.35} & 3.94 & 11.69 & 6.16 \\
     & Genre init. & 0.82 & 0.63 & 1.80 & 0.98 & 0.72 & 0.56 & 1.62 & 0.87 & \underline{5.35} & 3.94 & 11.69 & 6.16 \\
     & Best correction & 1.04 & 0.81 & 2.24 & 1.23 & 0.99 & 0.71 & 2.69 & 1.30 & 5.18 & 3.82 & 11.99 & 6.21 \\
     & \textbf{WeightRec (Ours)} & \textbf{1.90}$^{*}$ & \textbf{1.40}$^{*}$ & 4.50$^{*}$ & \textbf{2.31}$^{*}$ & \textbf{1.43}$^{*}$ & \textbf{1.05}$^{*}$ & \textbf{3.87}$^{*}$ & \textbf{1.89}$^{*}$ & \textbf{5.79}$^{*}$ & \textbf{4.34}$^{*}$ & \textbf{13.44}$^{*}$ & \textbf{7.04}$^{*}$ \\
    \midrule
    \multirow{8}{*}{\rotatebox{90}{BERT4Rec}} & Unedited & 0.98 & 0.71 & 2.28 & 1.16 & 0.92 & 0.72 & 2.37 & 1.24 & 5.01 & 3.78 & 12.82 & 6.52 \\
     & Fine-tuning$^{\dagger}$ & 0.99 & 0.75 & 3.34 & 1.58 & 1.16 & 0.90 & 2.52 & 1.37 & \textbf{6.12} & \textbf{4.52} & \textbf{14.86} & \textbf{7.60} \\
     & Prior shift & 0.85 & 0.64 & 1.98 & 1.04 & 0.58 & 0.44 & 2.23 & 1.00 & 4.99 & 3.79 & 12.86 & 6.54 \\
     & PDA & 0.85 & 0.63 & 2.36 & 1.15 & 1.07 & 0.82 & 2.63 & 1.36 & 5.65 & 4.23 & 13.54 & 7.01 \\
     & Folding-in & \underline{1.24} & \underline{0.85} & \underline{3.68} & \underline{1.69} & \underline{1.30} & \underline{0.94} & \underline{3.87} & \underline{1.83} & 5.01 & 3.78 & 12.82 & 6.52 \\
     & Genre init. & 0.94 & 0.69 & 2.19 & 1.12 & 0.89 & 0.70 & 2.34 & 1.22 & 5.01 & 3.78 & 12.82 & 6.52 \\
     & Best correction & 0.85 & 0.63 & 2.36 & 1.15 & 1.14 & 0.86 & 3.01 & 1.51 & 5.65 & 4.23 & 13.54 & 7.01 \\
     & \textbf{WeightRec (Ours)} & \textbf{1.48}$^{*}$ & \textbf{1.04}$^{*}$ & \textbf{4.02}$^{*}$ & \textbf{1.92}$^{*}$ & \textbf{1.71}$^{*}$ & \textbf{1.30}$^{*}$ & \textbf{4.83}$^{*}$ & \textbf{2.37}$^{*}$ & \underline{5.66} & \underline{4.29} & \underline{13.77} & \underline{7.11} \\
    \bottomrule
  \end{tabular*}
  \vspace{-2mm}
\end{table*}

%% file: chapters/tab_stream.tex
\begin{table*}[t]
  \centering
  \caption{Test accuracy (\%) after three streaming update steps, averaged over three seeds. Each step receives a new batch of feedback. Incremental SFT fine-tunes the model from the previous step on the new batch, while Cumulative SFT fine-tunes the original trained model on all feedback received so far. WeightRec edits the trained model with an edit chain re-materialized on all arrived feedback, and SFT + WeightRec applies the same kind of chain on top of the Incremental SFT model at each step. Within each column, the best result is in bold and the second best is underlined. $^{*}$The paired 95\% confidence interval of the method minus Incremental SFT lies above zero.}
  \label{tab:stream}
  \vspace{-3mm}
  \small
  \setlength{\tabcolsep}{2.2pt}
  \begin{tabular*}{\textwidth}{@{\extracolsep{\fill}}ll|ccc|ccc|ccc@{}}
    \toprule
    & & \multicolumn{3}{c|}{\textbf{Beauty}} & \multicolumn{3}{c|}{\textbf{Toys}} & \multicolumn{3}{c}{\textbf{MovieLens-1M}} \\
    \textbf{Metric} & \textbf{Method} & SASRec & GRU4Rec & BERT4Rec & SASRec & GRU4Rec & BERT4Rec & SASRec & GRU4Rec & BERT4Rec \\
    \midrule
    \multirow{5}{*}{N@10} & No update & 1.32 & 1.07 & 1.16 & 1.33 & 0.95 & 1.24 & 6.63 & 6.16 & 6.52 \\
     & Incremental SFT & 1.77 & 1.59 & 1.72 & 1.42 & 1.15 & 1.54 & 7.19 & 6.28 & \underline{7.32} \\
     & Cumulative SFT & 1.79 & \underline{2.23} & 1.81 & 1.49 & 1.34 & 1.54 & 6.97 & 6.45 & \textbf{7.47} \\
     & \textbf{WeightRec} & \underline{2.22}$^{*}$ & 1.96$^{*}$ & \underline{2.20}$^{*}$ & \textbf{2.12}$^{*}$ & \textbf{1.87}$^{*}$ & \underline{1.99}$^{*}$ & \underline{7.29} & \textbf{6.94}$^{*}$ & 7.07 \\
     & \textbf{SFT + WeightRec} & \textbf{2.42}$^{*}$ & \textbf{2.37}$^{*}$ & \textbf{2.85}$^{*}$ & \underline{1.75}$^{*}$ & \underline{1.54}$^{*}$ & \textbf{2.09}$^{*}$ & \textbf{7.31} & \underline{6.57} & 7.19 \\
    \midrule
    \multirow{5}{*}{H@10} & No update & 2.51 & 1.95 & 2.28 & 2.50 & 1.76 & 2.37 & 12.97 & 11.69 & 12.82 \\
     & Incremental SFT & 3.53 & 3.08 & 3.51 & 2.56 & 2.19 & 2.99 & \underline{14.32} & 12.15 & \underline{14.28} \\
     & Cumulative SFT & 3.52 & \underline{4.42} & 3.70 & 2.72 & 2.65 & 2.85 & 13.84 & 12.49 & \textbf{14.56} \\
     & \textbf{WeightRec} & \underline{4.33}$^{*}$ & 3.92$^{*}$ & \underline{4.84}$^{*}$ & \textbf{4.23}$^{*}$ & \textbf{3.94}$^{*}$ & \underline{4.12}$^{*}$ & 13.88 & \textbf{13.28}$^{*}$ & 13.85 \\
     & \textbf{SFT + WeightRec} & \textbf{4.93}$^{*}$ & \textbf{4.94}$^{*}$ & \textbf{6.00}$^{*}$ & \underline{3.67}$^{*}$ & \underline{3.16}$^{*}$ & \textbf{4.33}$^{*}$ & \textbf{14.43} & \underline{12.60} & 14.01 \\
    \bottomrule
  \end{tabular*}
  \vspace{-2mm}
\end{table*}

%% file: chapters/tab_ablation.tex
\begin{table}[t]
  \centering
  \caption{Contribution of each component (\%, mean over three seeds). Each round spends 20 edit ideas with at most 80 evaluations. The best result in each column is in bold and the second best is underlined.}
  \label{tab:ablation}
  \vspace{-3mm}
  \small
  \setlength{\tabcolsep}{2.4pt}
  \begin{tabular}{@{}l|cc|cc@{}}
    \toprule
    & \multicolumn{2}{c|}{\textbf{Beauty / GRU4Rec}} & \multicolumn{2}{c}{\textbf{ML-1M / GRU4Rec}} \\
    \textbf{Variant} & \textbf{N@10} & \textbf{H@10} & \textbf{N@10} & \textbf{H@10} \\
    \midrule
    Unedited & 1.07 & 1.95 & 6.16 & 11.69 \\
    \midrule
    \textbf{WeightRec-Evo} & \underline{2.31} & \underline{4.50} & \textbf{7.04} & \textbf{13.44} \\
    \quad w/o LLM (fixed editor) & 1.28 & 2.36 & \underline{6.80} & \underline{12.92} \\
    \midrule
    \textbf{WeightRec-Lite} & \textbf{2.42} & \textbf{4.88} & 6.72 & 12.81 \\
    \quad w/o forward validation & 1.99 & 3.78 & 5.31 & 10.38 \\
    \quad w/o re-materialization & 1.57 & 3.17 & 6.58 & 12.55 \\
    \quad w/o idea budget (20 evals) & 1.64 & 3.01 & 6.43 & 12.52 \\
    \quad w/o LLM (20-eval search) & 1.47 & 2.96 & 6.24 & 12.01 \\
    \quad w/o LLM (100-eval search) & 1.63 & 3.21 & 6.37 & 12.39 \\
    \bottomrule
  \end{tabular}
  \vspace{-3mm}
\end{table}

%% file: chapters/5_related_work.tex
\section{Related Work}
\label{sec:related}

\textbf{Correcting trained recommenders.}
A large body of work adjusts recommenders for biases in the observed data, through inverse propensity weighting~\cite{IPS}, causal adjustment of popularity during or after training~\cite{PDA,MACR}, and other debiasing schemes surveyed by \citet{BiasSurvey}.
Prior shift corrections from classification adjust the logits for a change in class frequencies~\cite{PriorShift,LogitAdjust}.
Cold-start methods produce embeddings for new items from content or from meta-learned warm-up~\cite{DropoutNet,MeLU,MWUF}, and folding-in places a new item at the mean of the users who chose it~\cite{FoldingIn}.
Streaming recommenders are kept up to date by incremental or meta-learned fine-tuning on each new batch of interactions~\cite{Rendle2008,SML}.
These methods each target one cause or update every weight by gradients, whereas WeightRec locates the cause by diagnosing the weights and writes the update as a program, which can replace a gradient update or run on top of it.

\textbf{Model editing.}
In language models, editing methods change specific behaviors by locating and rewriting weights~\cite{ROME,MEMIT}, by training hypernetworks that predict weight updates~\cite{MEND}, or by adding task vectors~\cite{TaskArithmetic}, as surveyed by \citet{EditSurvey}.
Recommendation unlearning removes the influence of selected interactions~\cite{RecUnlearning}, and recommendation editing has been formulated as correcting specific unwanted recommendations of a trained model~\cite{RecEditing}.
Causal tracing locates the weights behind a known behavior by intervening on hidden states and observing the change in the output~\cite{ROME}.
These approaches start from a specified target behavior whose change is local and can be verified prompt by prompt, while WeightRec targets the aggregate accuracy for future users under edits that move every top-$K$ list, so it locates with a similar loop but decides by forward-validated aggregate gains.

\textbf{Post-training, reinforcement learning, and self-evolution.}
Post-training adapts a model by gradient steps on supervised examples or on rewards from rollouts~\cite{InstructGPT,PPO,DeepSeekMath}, which updates every weight from the data it sees~\cite{Kirkpatrick2017}, while self-evolving agents improve the prompts, memory, or tools around a fixed model~\cite{Reflexion,Voyager,SelfEvolveSurvey}.
WeightRec differs from both, since an agent outside the model writes the update as a program on its weights, which changes the deployed model itself and can be read and reverted.

\textbf{LLMs for recommendation and program search.}
LLMs have served as recommenders through instruction tuning and prompting~\cite{P5,TALLRec}, as data augmenters~\cite{LLMRec}, and as agents that call recommendation tools~\cite{RecMind,InteRecAgent}, as reviewed by \citet{LLMRecSurvey}.
Separately, LLMs optimize prompts and programs from scored feedback~\cite{OPRO,FunSearch,AlphaEvolve}, although reusing a fixed holdout to judge many adaptive candidates over-fits it~\cite{ReusableHoldout}.
WeightRec lets an LLM update the weights of a conventional recommender and scores its programs forward in time.

%% file: chapters/6_conclusion.tex
\section{Conclusion}
\label{sec:conclusion}

We presented WeightRec, an update paradigm in which an LLM updates a trained recommender by locating what limits it, writing the update as a symbolic program over feedback statistics, and keeping it only if it gains on later feedback.
Since each update is a program, the accepted chain was re-materialized as feedback arrived, which let WeightRec update a recommender alone or as a plug-in on top of fine-tuning.
Experiments on three datasets and three backbones showed that both uses outperformed SFT in streaming updates in most settings, and never fell significantly below it.

%% file: main.bbl

\begin{thebibliography}{45}


\ifx \showCODEN    \undefined \def \showCODEN     #1{\unskip}     \fi
\ifx \showISBNx    \undefined \def \showISBNx     #1{\unskip}     \fi
\ifx \showISBNxiii \undefined \def \showISBNxiii  #1{\unskip}     \fi
\ifx \showISSN     \undefined \def \showISSN      #1{\unskip}     \fi
\ifx \showLCCN     \undefined \def \showLCCN      #1{\unskip}     \fi
\ifx \shownote     \undefined \def \shownote      #1{#1}          \fi
\ifx \showarticletitle \undefined \def \showarticletitle #1{#1}   \fi
\ifx \showURL      \undefined \def \showURL       {\relax}        \fi
\providecommand\bibfield[2]{#2}
\providecommand\bibinfo[2]{#2}
\providecommand\natexlab[1]{#1}
\providecommand\showeprint[2][]{arXiv:#2}

\bibitem[Bao et~al\mbox{.}(2023)]%
        {TALLRec}
\bibfield{author}{\bibinfo{person}{Keqin Bao}, \bibinfo{person}{Jizhi Zhang},
  \bibinfo{person}{Yang Zhang}, \bibinfo{person}{Wenjie Wang},
  \bibinfo{person}{Fuli Feng}, {and} \bibinfo{person}{Xiangnan He}.}
  \bibinfo{year}{2023}\natexlab{}.
\newblock \showarticletitle{TALLRec: An Effective and Efficient Tuning
  Framework to Align Large Language Model with Recommendation}. In
  \bibinfo{booktitle}{\emph{RecSys}}. \bibinfo{publisher}{{ACM}},
  \bibinfo{pages}{1007--1014}.
\newblock


\bibitem[Chen et~al\mbox{.}(2022)]%
        {RecUnlearning}
\bibfield{author}{\bibinfo{person}{Chong Chen}, \bibinfo{person}{Fei Sun},
  \bibinfo{person}{Min Zhang}, {and} \bibinfo{person}{Bolin Ding}.}
  \bibinfo{year}{2022}\natexlab{}.
\newblock \showarticletitle{Recommendation Unlearning}. In
  \bibinfo{booktitle}{\emph{{WWW}}}. \bibinfo{publisher}{{ACM}},
  \bibinfo{pages}{2768--2777}.
\newblock


\bibitem[Chen et~al\mbox{.}(2023)]%
        {BiasSurvey}
\bibfield{author}{\bibinfo{person}{Jiawei Chen}, \bibinfo{person}{Hande Dong},
  \bibinfo{person}{Xiang Wang}, \bibinfo{person}{Fuli Feng},
  \bibinfo{person}{Meng Wang}, {and} \bibinfo{person}{Xiangnan He}.}
  \bibinfo{year}{2023}\natexlab{}.
\newblock \showarticletitle{Bias and Debias in Recommender System: {A} Survey
  and Future Directions}.
\newblock \bibinfo{journal}{\emph{{ACM} Trans. Inf. Syst.}}
  \bibinfo{volume}{41}, \bibinfo{number}{3} (\bibinfo{year}{2023}),
  \bibinfo{pages}{67:1--67:39}.
\newblock


\bibitem[Dwork et~al\mbox{.}(2015)]%
        {ReusableHoldout}
\bibfield{author}{\bibinfo{person}{Cynthia Dwork}, \bibinfo{person}{Vitaly
  Feldman}, \bibinfo{person}{Moritz Hardt}, \bibinfo{person}{Toniann Pitassi},
  \bibinfo{person}{Omer Reingold}, {and} \bibinfo{person}{Aaron Roth}.}
  \bibinfo{year}{2015}\natexlab{}.
\newblock \showarticletitle{The reusable holdout: Preserving validity in
  adaptive data analysis}.
\newblock \bibinfo{journal}{\emph{Science}} \bibinfo{volume}{349},
  \bibinfo{number}{6248} (\bibinfo{year}{2015}), \bibinfo{pages}{636--638}.
\newblock


\bibitem[Gao et~al\mbox{.}(2026)]%
        {SelfEvolveSurvey}
\bibfield{author}{\bibinfo{person}{Huan{-}ang Gao}, \bibinfo{person}{Jiayi
  Geng}, \bibinfo{person}{Wenyue Hua}, \bibinfo{person}{Mengkang Hu},
  \bibinfo{person}{Xinzhe Juan}, \bibinfo{person}{Hongzhang Liu},
  \bibinfo{person}{Shilong Liu}, \bibinfo{person}{Jiahao Qiu},
  \bibinfo{person}{Xuan Qi}, \bibinfo{person}{Qihan Ren},
  \bibinfo{person}{Yiran Wu}, \bibinfo{person}{Hongru Wang},
  \bibinfo{person}{Han Xiao}, \bibinfo{person}{Yuhang Zhou},
  \bibinfo{person}{Shaokun Zhang}, \bibinfo{person}{Jiayi Zhang},
  \bibinfo{person}{Jinyu Xiang}, \bibinfo{person}{Yixiong Fang},
  \bibinfo{person}{Qiwen Zhao}, \bibinfo{person}{Dongrui Liu},
  \bibinfo{person}{Cheng Qian}, \bibinfo{person}{Zhenhailong Wang},
  \bibinfo{person}{Minda Hu}, \bibinfo{person}{Huazheng Wang},
  \bibinfo{person}{Qingyun Wu}, \bibinfo{person}{Heng Ji}, {and}
  \bibinfo{person}{Mengdi Wang}.} \bibinfo{year}{2026}\natexlab{}.
\newblock \showarticletitle{A Survey of Self-Evolving Agents: What, When, How,
  and Where to Evolve on the Path to Artificial Super Intelligence}.
\newblock \bibinfo{journal}{\emph{Trans. Mach. Learn. Res.}}
  \bibinfo{volume}{2026} (\bibinfo{year}{2026}).
\newblock


\bibitem[Geng et~al\mbox{.}(2022)]%
        {P5}
\bibfield{author}{\bibinfo{person}{Shijie Geng}, \bibinfo{person}{Shuchang
  Liu}, \bibinfo{person}{Zuohui Fu}, \bibinfo{person}{Yingqiang Ge}, {and}
  \bibinfo{person}{Yongfeng Zhang}.} \bibinfo{year}{2022}\natexlab{}.
\newblock \showarticletitle{Recommendation as Language Processing {(RLP):} {A}
  Unified Pretrain, Personalized Prompt {\&} Predict Paradigm {(P5)}}. In
  \bibinfo{booktitle}{\emph{RecSys}}. \bibinfo{publisher}{{ACM}},
  \bibinfo{pages}{299--315}.
\newblock


\bibitem[Harper and Konstan(2016)]%
        {MovieLens}
\bibfield{author}{\bibinfo{person}{F.~Maxwell Harper} {and}
  \bibinfo{person}{Joseph~A. Konstan}.} \bibinfo{year}{2016}\natexlab{}.
\newblock \showarticletitle{The MovieLens Datasets: History and Context}.
\newblock \bibinfo{journal}{\emph{{ACM} Trans. Interact. Intell. Syst.}}
  \bibinfo{volume}{5}, \bibinfo{number}{4} (\bibinfo{year}{2016}),
  \bibinfo{pages}{19:1--19:19}.
\newblock


\bibitem[Hidasi et~al\mbox{.}(2016)]%
        {GRU4Rec}
\bibfield{author}{\bibinfo{person}{Bal{\'{a}}zs Hidasi},
  \bibinfo{person}{Alexandros Karatzoglou}, \bibinfo{person}{Linas Baltrunas},
  {and} \bibinfo{person}{Domonkos Tikk}.} \bibinfo{year}{2016}\natexlab{}.
\newblock \showarticletitle{Session-based Recommendations with Recurrent Neural
  Networks}. In \bibinfo{booktitle}{\emph{{ICLR} (Poster)}}.
\newblock


\bibitem[Huang et~al\mbox{.}(2025)]%
        {InteRecAgent}
\bibfield{author}{\bibinfo{person}{Xu Huang}, \bibinfo{person}{Jianxun Lian},
  \bibinfo{person}{Yuxuan Lei}, \bibinfo{person}{Jing Yao},
  \bibinfo{person}{Defu Lian}, {and} \bibinfo{person}{Xing Xie}.}
  \bibinfo{year}{2025}\natexlab{}.
\newblock \showarticletitle{Recommender {AI} Agent: Integrating Large Language
  Models for Interactive Recommendations}.
\newblock \bibinfo{journal}{\emph{{ACM} Trans. Inf. Syst.}}
  \bibinfo{volume}{43}, \bibinfo{number}{4} (\bibinfo{year}{2025}),
  \bibinfo{pages}{96:1--96:33}.
\newblock


\bibitem[Ilharco et~al\mbox{.}(2023)]%
        {TaskArithmetic}
\bibfield{author}{\bibinfo{person}{Gabriel Ilharco},
  \bibinfo{person}{Marco~T{\'{u}}lio Ribeiro}, \bibinfo{person}{Mitchell
  Wortsman}, \bibinfo{person}{Ludwig Schmidt}, \bibinfo{person}{Hannaneh
  Hajishirzi}, {and} \bibinfo{person}{Ali Farhadi}.}
  \bibinfo{year}{2023}\natexlab{}.
\newblock \showarticletitle{Editing models with task arithmetic}. In
  \bibinfo{booktitle}{\emph{{ICLR}}}. \bibinfo{publisher}{OpenReview.net}.
\newblock


\bibitem[Ji et~al\mbox{.}(2023)]%
        {RecLeakage}
\bibfield{author}{\bibinfo{person}{Yitong Ji}, \bibinfo{person}{Aixin Sun},
  \bibinfo{person}{Jie Zhang}, {and} \bibinfo{person}{Chenliang Li}.}
  \bibinfo{year}{2023}\natexlab{}.
\newblock \showarticletitle{A Critical Study on Data Leakage in Recommender
  System Offline Evaluation}.
\newblock \bibinfo{journal}{\emph{{ACM} Trans. Inf. Syst.}}
  \bibinfo{volume}{41}, \bibinfo{number}{3} (\bibinfo{year}{2023}),
  \bibinfo{pages}{75:1--75:27}.
\newblock


\bibitem[Kang and McAuley(2018)]%
        {SASRec}
\bibfield{author}{\bibinfo{person}{Wang{-}Cheng Kang} {and}
  \bibinfo{person}{Julian~J. McAuley}.} \bibinfo{year}{2018}\natexlab{}.
\newblock \showarticletitle{Self-Attentive Sequential Recommendation}. In
  \bibinfo{booktitle}{\emph{{ICDM}}}. \bibinfo{publisher}{{IEEE} Computer
  Society}, \bibinfo{pages}{197--206}.
\newblock


\bibitem[Kirkpatrick et~al\mbox{.}(2016)]%
        {Kirkpatrick2017}
\bibfield{author}{\bibinfo{person}{James Kirkpatrick}, \bibinfo{person}{Razvan
  Pascanu}, \bibinfo{person}{Neil~C. Rabinowitz}, \bibinfo{person}{Joel
  Veness}, \bibinfo{person}{Guillaume Desjardins}, \bibinfo{person}{Andrei~A.
  Rusu}, \bibinfo{person}{Kieran Milan}, \bibinfo{person}{John Quan},
  \bibinfo{person}{Tiago Ramalho}, \bibinfo{person}{Agnieszka
  Grabska{-}Barwinska}, \bibinfo{person}{Demis Hassabis},
  \bibinfo{person}{Claudia Clopath}, \bibinfo{person}{Dharshan Kumaran}, {and}
  \bibinfo{person}{Raia Hadsell}.} \bibinfo{year}{2016}\natexlab{}.
\newblock \showarticletitle{Overcoming catastrophic forgetting in neural
  networks}.
\newblock \bibinfo{journal}{\emph{CoRR}}  \bibinfo{volume}{abs/1612.00796}
  (\bibinfo{year}{2016}).
\newblock


\bibitem[Krichene and Rendle(2020)]%
        {SampledMetrics}
\bibfield{author}{\bibinfo{person}{Walid Krichene} {and}
  \bibinfo{person}{Steffen Rendle}.} \bibinfo{year}{2020}\natexlab{}.
\newblock \showarticletitle{On Sampled Metrics for Item Recommendation}. In
  \bibinfo{booktitle}{\emph{{KDD}}}. \bibinfo{publisher}{{ACM}},
  \bibinfo{pages}{1748--1757}.
\newblock


\bibitem[Lai et~al\mbox{.}(2024)]%
        {RecEditing}
\bibfield{author}{\bibinfo{person}{Chengyu Lai}, \bibinfo{person}{Sheng Zhou},
  \bibinfo{person}{Zhimeng Jiang}, \bibinfo{person}{Qiaoyu Tan},
  \bibinfo{person}{Yuanchen Bei}, \bibinfo{person}{Jiawei Chen},
  \bibinfo{person}{Ningyu Zhang}, {and} \bibinfo{person}{Jiajun Bu}.}
  \bibinfo{year}{2024}\natexlab{}.
\newblock \showarticletitle{Better Late Than Never: Formulating and
  Benchmarking Recommendation Editing}.
\newblock \bibinfo{journal}{\emph{CoRR}}  \bibinfo{volume}{abs/2406.04553}
  (\bibinfo{year}{2024}).
\newblock


\bibitem[Lee et~al\mbox{.}(2019)]%
        {MeLU}
\bibfield{author}{\bibinfo{person}{Hoyeop Lee}, \bibinfo{person}{Jinbae Im},
  \bibinfo{person}{Seongwon Jang}, \bibinfo{person}{Hyunsouk Cho}, {and}
  \bibinfo{person}{Sehee Chung}.} \bibinfo{year}{2019}\natexlab{}.
\newblock \showarticletitle{MeLU: Meta-Learned User Preference Estimator for
  Cold-Start Recommendation}. In \bibinfo{booktitle}{\emph{{KDD}}}.
  \bibinfo{publisher}{{ACM}}, \bibinfo{pages}{1073--1082}.
\newblock


\bibitem[McAuley et~al\mbox{.}(2015)]%
        {AmazonData}
\bibfield{author}{\bibinfo{person}{Julian~J. McAuley},
  \bibinfo{person}{Christopher Targett}, \bibinfo{person}{Qinfeng Shi}, {and}
  \bibinfo{person}{Anton van~den Hengel}.} \bibinfo{year}{2015}\natexlab{}.
\newblock \showarticletitle{Image-Based Recommendations on Styles and
  Substitutes}. In \bibinfo{booktitle}{\emph{{SIGIR}}}.
  \bibinfo{publisher}{{ACM}}, \bibinfo{pages}{43--52}.
\newblock


\bibitem[Meng et~al\mbox{.}(2022)]%
        {ROME}
\bibfield{author}{\bibinfo{person}{Kevin Meng}, \bibinfo{person}{David Bau},
  \bibinfo{person}{Alex Andonian}, {and} \bibinfo{person}{Yonatan Belinkov}.}
  \bibinfo{year}{2022}\natexlab{}.
\newblock \showarticletitle{Locating and Editing Factual Associations in
  {GPT}}. In \bibinfo{booktitle}{\emph{NeurIPS}}.
\newblock


\bibitem[Meng et~al\mbox{.}(2023)]%
        {MEMIT}
\bibfield{author}{\bibinfo{person}{Kevin Meng}, \bibinfo{person}{Arnab~Sen
  Sharma}, \bibinfo{person}{Alex~J. Andonian}, \bibinfo{person}{Yonatan
  Belinkov}, {and} \bibinfo{person}{David Bau}.}
  \bibinfo{year}{2023}\natexlab{}.
\newblock \showarticletitle{Mass-Editing Memory in a Transformer}. In
  \bibinfo{booktitle}{\emph{{ICLR}}}. \bibinfo{publisher}{OpenReview.net}.
\newblock


\bibitem[Menon et~al\mbox{.}(2021)]%
        {LogitAdjust}
\bibfield{author}{\bibinfo{person}{Aditya~Krishna Menon},
  \bibinfo{person}{Sadeep Jayasumana}, \bibinfo{person}{Ankit~Singh Rawat},
  \bibinfo{person}{Himanshu Jain}, \bibinfo{person}{Andreas Veit}, {and}
  \bibinfo{person}{Sanjiv Kumar}.} \bibinfo{year}{2021}\natexlab{}.
\newblock \showarticletitle{Long-tail learning via logit adjustment}. In
  \bibinfo{booktitle}{\emph{{ICLR}}}. \bibinfo{publisher}{OpenReview.net}.
\newblock


\bibitem[Mitchell et~al\mbox{.}(2022)]%
        {MEND}
\bibfield{author}{\bibinfo{person}{Eric Mitchell}, \bibinfo{person}{Charles
  Lin}, \bibinfo{person}{Antoine Bosselut}, \bibinfo{person}{Chelsea Finn},
  {and} \bibinfo{person}{Christopher~D. Manning}.}
  \bibinfo{year}{2022}\natexlab{}.
\newblock \showarticletitle{Fast Model Editing at Scale}. In
  \bibinfo{booktitle}{\emph{{ICLR}}}. \bibinfo{publisher}{OpenReview.net}.
\newblock


\bibitem[Ni et~al\mbox{.}(2019)]%
        {Amazon2018}
\bibfield{author}{\bibinfo{person}{Jianmo Ni}, \bibinfo{person}{Jiacheng Li},
  {and} \bibinfo{person}{Julian~J. McAuley}.} \bibinfo{year}{2019}\natexlab{}.
\newblock \showarticletitle{Justifying Recommendations using Distantly-Labeled
  Reviews and Fine-Grained Aspects}. In
  \bibinfo{booktitle}{\emph{{EMNLP/IJCNLP} {(1)}}}.
  \bibinfo{publisher}{Association for Computational Linguistics},
  \bibinfo{pages}{188--197}.
\newblock


\bibitem[Novikov et~al\mbox{.}(2025)]%
        {AlphaEvolve}
\bibfield{author}{\bibinfo{person}{Alexander Novikov},
  \bibinfo{person}{Ng{\^{a}}n Vu}, \bibinfo{person}{Marvin Eisenberger},
  \bibinfo{person}{Emilien Dupont}, \bibinfo{person}{Po{-}Sen Huang},
  \bibinfo{person}{Adam~Zsolt Wagner}, \bibinfo{person}{Sergey Shirobokov},
  \bibinfo{person}{Borislav Kozlovskii}, \bibinfo{person}{Francisco J.~R.
  Ruiz}, \bibinfo{person}{Abbas Mehrabian}, \bibinfo{person}{M.~Pawan Kumar},
  \bibinfo{person}{Abigail See}, \bibinfo{person}{Swarat Chaudhuri},
  \bibinfo{person}{George Holland}, \bibinfo{person}{Alex Davies},
  \bibinfo{person}{Sebastian Nowozin}, \bibinfo{person}{Pushmeet Kohli}, {and}
  \bibinfo{person}{Matej Balog}.} \bibinfo{year}{2025}\natexlab{}.
\newblock \showarticletitle{AlphaEvolve: {A} coding agent for scientific and
  algorithmic discovery}.
\newblock \bibinfo{journal}{\emph{CoRR}}  \bibinfo{volume}{abs/2506.13131}
  (\bibinfo{year}{2025}).
\newblock


\bibitem[Ouyang et~al\mbox{.}(2022)]%
        {InstructGPT}
\bibfield{author}{\bibinfo{person}{Long Ouyang}, \bibinfo{person}{Jeffrey Wu},
  \bibinfo{person}{Xu Jiang}, \bibinfo{person}{Diogo Almeida},
  \bibinfo{person}{Carroll~L. Wainwright}, \bibinfo{person}{Pamela Mishkin},
  \bibinfo{person}{Chong Zhang}, \bibinfo{person}{Sandhini Agarwal},
  \bibinfo{person}{Katarina Slama}, \bibinfo{person}{Alex Ray},
  \bibinfo{person}{John Schulman}, \bibinfo{person}{Jacob Hilton},
  \bibinfo{person}{Fraser Kelton}, \bibinfo{person}{Luke Miller},
  \bibinfo{person}{Maddie Simens}, \bibinfo{person}{Amanda Askell},
  \bibinfo{person}{Peter Welinder}, \bibinfo{person}{Paul~F. Christiano},
  \bibinfo{person}{Jan Leike}, {and} \bibinfo{person}{Ryan Lowe}.}
  \bibinfo{year}{2022}\natexlab{}.
\newblock \showarticletitle{Training language models to follow instructions
  with human feedback}. In \bibinfo{booktitle}{\emph{NeurIPS}}.
\newblock


\bibitem[Rendle and Schmidt{-}Thieme(2008)]%
        {Rendle2008}
\bibfield{author}{\bibinfo{person}{Steffen Rendle} {and} \bibinfo{person}{Lars
  Schmidt{-}Thieme}.} \bibinfo{year}{2008}\natexlab{}.
\newblock \showarticletitle{Online-updating regularized kernel matrix
  factorization models for large-scale recommender systems}. In
  \bibinfo{booktitle}{\emph{RecSys}}. \bibinfo{publisher}{{ACM}},
  \bibinfo{pages}{251--258}.
\newblock


\bibitem[Romera{-}Paredes et~al\mbox{.}(2024)]%
        {FunSearch}
\bibfield{author}{\bibinfo{person}{Bernardino Romera{-}Paredes},
  \bibinfo{person}{Mohammadamin Barekatain}, \bibinfo{person}{Alexander
  Novikov}, \bibinfo{person}{Matej Balog}, \bibinfo{person}{M.~Pawan Kumar},
  \bibinfo{person}{Emilien Dupont}, \bibinfo{person}{Francisco J.~R. Ruiz},
  \bibinfo{person}{Jordan~S. Ellenberg}, \bibinfo{person}{Pengming Wang},
  \bibinfo{person}{Omar Fawzi}, \bibinfo{person}{Pushmeet Kohli}, {and}
  \bibinfo{person}{Alhussein Fawzi}.} \bibinfo{year}{2024}\natexlab{}.
\newblock \showarticletitle{Mathematical discoveries from program search with
  large language models}.
\newblock \bibinfo{journal}{\emph{Nat.}} \bibinfo{volume}{625},
  \bibinfo{number}{7995} (\bibinfo{year}{2024}), \bibinfo{pages}{468--475}.
\newblock


\bibitem[Saerens et~al\mbox{.}(2002)]%
        {PriorShift}
\bibfield{author}{\bibinfo{person}{Marco Saerens}, \bibinfo{person}{Patrice
  Latinne}, {and} \bibinfo{person}{Christine Decaestecker}.}
  \bibinfo{year}{2002}\natexlab{}.
\newblock \showarticletitle{Adjusting the Outputs of a Classifier to New a
  Priori Probabilities: {A} Simple Procedure}.
\newblock \bibinfo{journal}{\emph{Neural Comput.}} \bibinfo{volume}{14},
  \bibinfo{number}{1} (\bibinfo{year}{2002}), \bibinfo{pages}{21--41}.
\newblock


\bibitem[Sarwar et~al\mbox{.}(2002)]%
        {FoldingIn}
\bibfield{author}{\bibinfo{person}{Badrul Sarwar}, \bibinfo{person}{George
  Karypis}, \bibinfo{person}{Joseph Konstan}, {and} \bibinfo{person}{John
  Riedl}.} \bibinfo{year}{2002}\natexlab{}.
\newblock \showarticletitle{Incremental singular value decomposition algorithms
  for highly scalable recommender systems}. In \bibinfo{booktitle}{\emph{Fifth
  international conference on computer and information science}},
  Vol.~\bibinfo{volume}{1}. \bibinfo{pages}{27--8}.
\newblock


\bibitem[Schnabel et~al\mbox{.}(2016)]%
        {IPS}
\bibfield{author}{\bibinfo{person}{Tobias Schnabel}, \bibinfo{person}{Adith
  Swaminathan}, \bibinfo{person}{Ashudeep Singh}, \bibinfo{person}{Navin
  Chandak}, {and} \bibinfo{person}{Thorsten Joachims}.}
  \bibinfo{year}{2016}\natexlab{}.
\newblock \showarticletitle{Recommendations as Treatments: Debiasing Learning
  and Evaluation}. In \bibinfo{booktitle}{\emph{{ICML}}}
  \emph{(\bibinfo{series}{{JMLR} Workshop and Conference Proceedings},
  Vol.~\bibinfo{volume}{48})}. \bibinfo{publisher}{JMLR.org},
  \bibinfo{pages}{1670--1679}.
\newblock


\bibitem[Schulman et~al\mbox{.}(2017)]%
        {PPO}
\bibfield{author}{\bibinfo{person}{John Schulman}, \bibinfo{person}{Filip
  Wolski}, \bibinfo{person}{Prafulla Dhariwal}, \bibinfo{person}{Alec Radford},
  {and} \bibinfo{person}{Oleg Klimov}.} \bibinfo{year}{2017}\natexlab{}.
\newblock \showarticletitle{Proximal Policy Optimization Algorithms}.
\newblock \bibinfo{journal}{\emph{CoRR}}  \bibinfo{volume}{abs/1707.06347}
  (\bibinfo{year}{2017}).
\newblock


\bibitem[Shao et~al\mbox{.}(2024)]%
        {DeepSeekMath}
\bibfield{author}{\bibinfo{person}{Zhihong Shao}, \bibinfo{person}{Peiyi Wang},
  \bibinfo{person}{Qihao Zhu}, \bibinfo{person}{Runxin Xu},
  \bibinfo{person}{Junxiao Song}, \bibinfo{person}{Mingchuan Zhang},
  \bibinfo{person}{Y.~K. Li}, \bibinfo{person}{Y. Wu}, {and}
  \bibinfo{person}{Daya Guo}.} \bibinfo{year}{2024}\natexlab{}.
\newblock \showarticletitle{DeepSeekMath: Pushing the Limits of Mathematical
  Reasoning in Open Language Models}.
\newblock \bibinfo{journal}{\emph{CoRR}}  \bibinfo{volume}{abs/2402.03300}
  (\bibinfo{year}{2024}).
\newblock


\bibitem[Shinn et~al\mbox{.}(2023)]%
        {Reflexion}
\bibfield{author}{\bibinfo{person}{Noah Shinn}, \bibinfo{person}{Federico
  Cassano}, \bibinfo{person}{Ashwin Gopinath}, \bibinfo{person}{Karthik
  Narasimhan}, {and} \bibinfo{person}{Shunyu Yao}.}
  \bibinfo{year}{2023}\natexlab{}.
\newblock \showarticletitle{Reflexion: language agents with verbal
  reinforcement learning}. In \bibinfo{booktitle}{\emph{NeurIPS}}.
\newblock


\bibitem[Sun et~al\mbox{.}(2019)]%
        {BERT4Rec}
\bibfield{author}{\bibinfo{person}{Fei Sun}, \bibinfo{person}{Jun Liu},
  \bibinfo{person}{Jian Wu}, \bibinfo{person}{Changhua Pei},
  \bibinfo{person}{Xiao Lin}, \bibinfo{person}{Wenwu Ou}, {and}
  \bibinfo{person}{Peng Jiang}.} \bibinfo{year}{2019}\natexlab{}.
\newblock \showarticletitle{BERT4Rec: Sequential Recommendation with
  Bidirectional Encoder Representations from Transformer}. In
  \bibinfo{booktitle}{\emph{{CIKM}}}. \bibinfo{publisher}{{ACM}},
  \bibinfo{pages}{1441--1450}.
\newblock


\bibitem[Volkovs et~al\mbox{.}(2017)]%
        {DropoutNet}
\bibfield{author}{\bibinfo{person}{Maksims Volkovs}, \bibinfo{person}{Guangwei
  Yu}, {and} \bibinfo{person}{Tomi Poutanen}.} \bibinfo{year}{2017}\natexlab{}.
\newblock \showarticletitle{DropoutNet: Addressing Cold Start in Recommender
  Systems}. In \bibinfo{booktitle}{\emph{{NIPS}}}. \bibinfo{pages}{4957--4966}.
\newblock


\bibitem[Wang et~al\mbox{.}(2024b)]%
        {Voyager}
\bibfield{author}{\bibinfo{person}{Guanzhi Wang}, \bibinfo{person}{Yuqi Xie},
  \bibinfo{person}{Yunfan Jiang}, \bibinfo{person}{Ajay Mandlekar},
  \bibinfo{person}{Chaowei Xiao}, \bibinfo{person}{Yuke Zhu},
  \bibinfo{person}{Linxi Fan}, {and} \bibinfo{person}{Anima Anandkumar}.}
  \bibinfo{year}{2024}\natexlab{b}.
\newblock \showarticletitle{Voyager: An Open-Ended Embodied Agent with Large
  Language Models}.
\newblock \bibinfo{journal}{\emph{Trans. Mach. Learn. Res.}}
  \bibinfo{volume}{2024} (\bibinfo{year}{2024}).
\newblock


\bibitem[Wang et~al\mbox{.}(2024a)]%
        {RecMind}
\bibfield{author}{\bibinfo{person}{Yancheng Wang}, \bibinfo{person}{Ziyan
  Jiang}, \bibinfo{person}{Zheng Chen}, \bibinfo{person}{Fan Yang},
  \bibinfo{person}{Yingxue Zhou}, \bibinfo{person}{Eunah Cho},
  \bibinfo{person}{Xing Fan}, \bibinfo{person}{Yanbin Lu},
  \bibinfo{person}{Xiaojiang Huang}, {and} \bibinfo{person}{Yingzhen Yang}.}
  \bibinfo{year}{2024}\natexlab{a}.
\newblock \showarticletitle{RecMind: Large Language Model Powered Agent For
  Recommendation}. In \bibinfo{booktitle}{\emph{{NAACL-HLT} (Findings)}}
  \emph{(\bibinfo{series}{Findings of {ACL}}, Vol.~\bibinfo{volume}{{NAACL}
  2024})}. \bibinfo{publisher}{Association for Computational Linguistics},
  \bibinfo{pages}{4351--4364}.
\newblock


\bibitem[Wei et~al\mbox{.}(2021)]%
        {MACR}
\bibfield{author}{\bibinfo{person}{Tianxin Wei}, \bibinfo{person}{Fuli Feng},
  \bibinfo{person}{Jiawei Chen}, \bibinfo{person}{Ziwei Wu},
  \bibinfo{person}{Jinfeng Yi}, {and} \bibinfo{person}{Xiangnan He}.}
  \bibinfo{year}{2021}\natexlab{}.
\newblock \showarticletitle{Model-Agnostic Counterfactual Reasoning for
  Eliminating Popularity Bias in Recommender System}. In
  \bibinfo{booktitle}{\emph{{KDD}}}. \bibinfo{publisher}{{ACM}},
  \bibinfo{pages}{1791--1800}.
\newblock


\bibitem[Wei et~al\mbox{.}(2024)]%
        {LLMRec}
\bibfield{author}{\bibinfo{person}{Wei Wei}, \bibinfo{person}{Xubin Ren},
  \bibinfo{person}{Jiabin Tang}, \bibinfo{person}{Qinyong Wang},
  \bibinfo{person}{Lixin Su}, \bibinfo{person}{Suqi Cheng},
  \bibinfo{person}{Junfeng Wang}, \bibinfo{person}{Dawei Yin}, {and}
  \bibinfo{person}{Chao Huang}.} \bibinfo{year}{2024}\natexlab{}.
\newblock \showarticletitle{LLMRec: Large Language Models with Graph
  Augmentation for Recommendation}. In \bibinfo{booktitle}{\emph{{WSDM}}}.
  \bibinfo{publisher}{{ACM}}, \bibinfo{pages}{806--815}.
\newblock


\bibitem[Wu et~al\mbox{.}(2024)]%
        {LLMRecSurvey}
\bibfield{author}{\bibinfo{person}{Likang Wu}, \bibinfo{person}{Zhi Zheng},
  \bibinfo{person}{Zhaopeng Qiu}, \bibinfo{person}{Hao Wang},
  \bibinfo{person}{Hongchao Gu}, \bibinfo{person}{Tingjia Shen},
  \bibinfo{person}{Chuan Qin}, \bibinfo{person}{Chen Zhu},
  \bibinfo{person}{Hengshu Zhu}, \bibinfo{person}{Qi Liu}, \bibinfo{person}{Hui
  Xiong}, {and} \bibinfo{person}{Enhong Chen}.}
  \bibinfo{year}{2024}\natexlab{}.
\newblock \showarticletitle{A survey on large language models for
  recommendation}.
\newblock \bibinfo{journal}{\emph{World Wide Web {(WWW)}}}
  \bibinfo{volume}{27}, \bibinfo{number}{5} (\bibinfo{year}{2024}),
  \bibinfo{pages}{60}.
\newblock


\bibitem[Yang et~al\mbox{.}(2024)]%
        {OPRO}
\bibfield{author}{\bibinfo{person}{Chengrun Yang}, \bibinfo{person}{Xuezhi
  Wang}, \bibinfo{person}{Yifeng Lu}, \bibinfo{person}{Hanxiao Liu},
  \bibinfo{person}{Quoc~V. Le}, \bibinfo{person}{Denny Zhou}, {and}
  \bibinfo{person}{Xinyun Chen}.} \bibinfo{year}{2024}\natexlab{}.
\newblock \showarticletitle{Large Language Models as Optimizers}. In
  \bibinfo{booktitle}{\emph{{ICLR}}}. \bibinfo{publisher}{OpenReview.net}.
\newblock


\bibitem[Yao et~al\mbox{.}(2023b)]%
        {ReAct}
\bibfield{author}{\bibinfo{person}{Shunyu Yao}, \bibinfo{person}{Jeffrey Zhao},
  \bibinfo{person}{Dian Yu}, \bibinfo{person}{Nan Du}, \bibinfo{person}{Izhak
  Shafran}, \bibinfo{person}{Karthik~R. Narasimhan}, {and}
  \bibinfo{person}{Yuan Cao}.} \bibinfo{year}{2023}\natexlab{b}.
\newblock \showarticletitle{ReAct: Synergizing Reasoning and Acting in Language
  Models}. In \bibinfo{booktitle}{\emph{{ICLR}}}.
  \bibinfo{publisher}{OpenReview.net}.
\newblock


\bibitem[Yao et~al\mbox{.}(2023a)]%
        {EditSurvey}
\bibfield{author}{\bibinfo{person}{Yunzhi Yao}, \bibinfo{person}{Peng Wang},
  \bibinfo{person}{Bozhong Tian}, \bibinfo{person}{Siyuan Cheng},
  \bibinfo{person}{Zhoubo Li}, \bibinfo{person}{Shumin Deng},
  \bibinfo{person}{Huajun Chen}, {and} \bibinfo{person}{Ningyu Zhang}.}
  \bibinfo{year}{2023}\natexlab{a}.
\newblock \showarticletitle{Editing Large Language Models: Problems, Methods,
  and Opportunities}. In \bibinfo{booktitle}{\emph{{EMNLP}}}.
  \bibinfo{publisher}{Association for Computational Linguistics},
  \bibinfo{pages}{10222--10240}.
\newblock


\bibitem[Zhang et~al\mbox{.}(2021)]%
        {PDA}
\bibfield{author}{\bibinfo{person}{Yang Zhang}, \bibinfo{person}{Fuli Feng},
  \bibinfo{person}{Xiangnan He}, \bibinfo{person}{Tianxin Wei},
  \bibinfo{person}{Chonggang Song}, \bibinfo{person}{Guohui Ling}, {and}
  \bibinfo{person}{Yongdong Zhang}.} \bibinfo{year}{2021}\natexlab{}.
\newblock \showarticletitle{Causal Intervention for Leveraging Popularity Bias
  in Recommendation}. In \bibinfo{booktitle}{\emph{{SIGIR}}}.
  \bibinfo{publisher}{{ACM}}, \bibinfo{pages}{11--20}.
\newblock


\bibitem[Zhang et~al\mbox{.}(2020)]%
        {SML}
\bibfield{author}{\bibinfo{person}{Yang Zhang}, \bibinfo{person}{Fuli Feng},
  \bibinfo{person}{Chenxu Wang}, \bibinfo{person}{Xiangnan He},
  \bibinfo{person}{Meng Wang}, \bibinfo{person}{Yan Li}, {and}
  \bibinfo{person}{Yongdong Zhang}.} \bibinfo{year}{2020}\natexlab{}.
\newblock \showarticletitle{How to Retrain Recommender System?: {A} Sequential
  Meta-Learning Method}. In \bibinfo{booktitle}{\emph{{SIGIR}}}.
  \bibinfo{publisher}{{ACM}}, \bibinfo{pages}{1479--1488}.
\newblock


\bibitem[Zhu et~al\mbox{.}(2021)]%
        {MWUF}
\bibfield{author}{\bibinfo{person}{Yongchun Zhu}, \bibinfo{person}{Ruobing
  Xie}, \bibinfo{person}{Fuzhen Zhuang}, \bibinfo{person}{Kaikai Ge},
  \bibinfo{person}{Ying Sun}, \bibinfo{person}{Xu Zhang}, \bibinfo{person}{Leyu
  Lin}, {and} \bibinfo{person}{Juan Cao}.} \bibinfo{year}{2021}\natexlab{}.
\newblock \showarticletitle{Learning to Warm Up Cold Item Embeddings for
  Cold-start Recommendation with Meta Scaling and Shifting Networks}. In
  \bibinfo{booktitle}{\emph{{SIGIR}}}. \bibinfo{publisher}{{ACM}},
  \bibinfo{pages}{1167--1176}.
\newblock


\end{thebibliography}
